\documentclass[prb,twocolumn,groupedaddress,showpacs]{revtex4-1}
\usepackage{graphics}
\usepackage{graphicx}
\usepackage{subfigure}
\usepackage{amssymb}
\usepackage{color}
\usepackage{amsmath}

\newcounter{magicrownumbers}

\begin{document}
\title{Polymorphism in Bi-based perovskite oxides:
a first-principles study}

\author{Akansha Singh,$^{1,2}$
Viveka N.~Singh,$^{1}$
Enric Canadell,$^{3}$ 
Jorge \'{I}\~{n}iguez,$^{4}$
and Oswaldo  Di\'{e}guez$^{1,2}$}
   
\email[Corresponding author: ]{dieguez$@$tau.ac.il}.

\affiliation{\vspace{0.3cm}
$^1$Department of Materials Science and Engineering,
             Faculty of Engineering, Tel Aviv University,
             Tel Aviv 69978, Israel
             \vspace{0.3cm} \\
$^2$The Raymond and Beverly Sackler Center for Computational
             Molecular and Materials Science,
             Tel Aviv University,
             Tel Aviv 69978, Israel
             \vspace{0.3cm} \\
$^3$Institut de Ci\`{e}ncia de Materials de Barcelona (ICMAB-CSIC),
             Campus UAB,
             E-08193 Bellaterra, Spain
             \vspace{0.3cm} \\
$^4$Materials Research and Technology Department,
             Luxembourg Institute of Science and Technology,
             5 avenue des Hauts-Fourneaux, L-4362 Esch/Alzette, Luxembourg
}

\begin{abstract}
Under normal conditions,
bulk crystals of BiScO$_3$, BiCrO$_3$, BiMnO$_3$, BiFeO$_3$, and
BiCoO$_3$ present three very different variations of the perovskite 
structure: an antipolar phase, a rhombohedral phase with a large 
polarization along the space diagonal of the pseudocubic unit cell, and
a supertetragonal phase with even larger polarization.
With the aim of understanding the causes for this variety,
we have used a genetic algorithm to search for minima in the
surface energy of these materials.
Our results show that the number of these minima is very large when compared
to that of typical ferroelectric perovskites like BaTiO$_3$ and PbTiO$_3$,
and that a fine energy balance between them results in the large structural
differences seen.
As byproducts of our search we have identified charge-ordering 
structures with low energy in BiMnO$_3$, and several phases with energies
that are similar to that of the ground state of BiCrO$_3$.
We have also found that a inverse supertetragonal phase exists in bulk,
likely to be favored in films epitaxially grown at large values of tensile
misfit strain.
\end{abstract}


\maketitle
\section{Introduction}
\label{sec:intro}

Thousands of research articles about BiFeO$_3$ have
been published since Wang {\em et al.}\cite{Wang2003S}
reported that epitaxial films of this material were multiferroic at room
temperature
(Ref.~\onlinecite{Catalan2009AM} reviews the properties of this oxide).
This effort has unvealed potential applications based on the 
possibility to control the electric polarization with a magnetic 
field\cite{Hur2004N} and the magnetization with an electric
field.\cite{Chu2008NM}
Other phenomena such as domain wall conductivity,\cite{Seidel2009NM}
novel photovoltaic effects,\cite{Yang2010NN} and the presence of a feature
akin to a morphotropic phase boundary in thin films\cite{Zeches2009S} 
have further fuelled research on BiFeO$_3$.

\begin{figure}
\includegraphics[width=80mm]{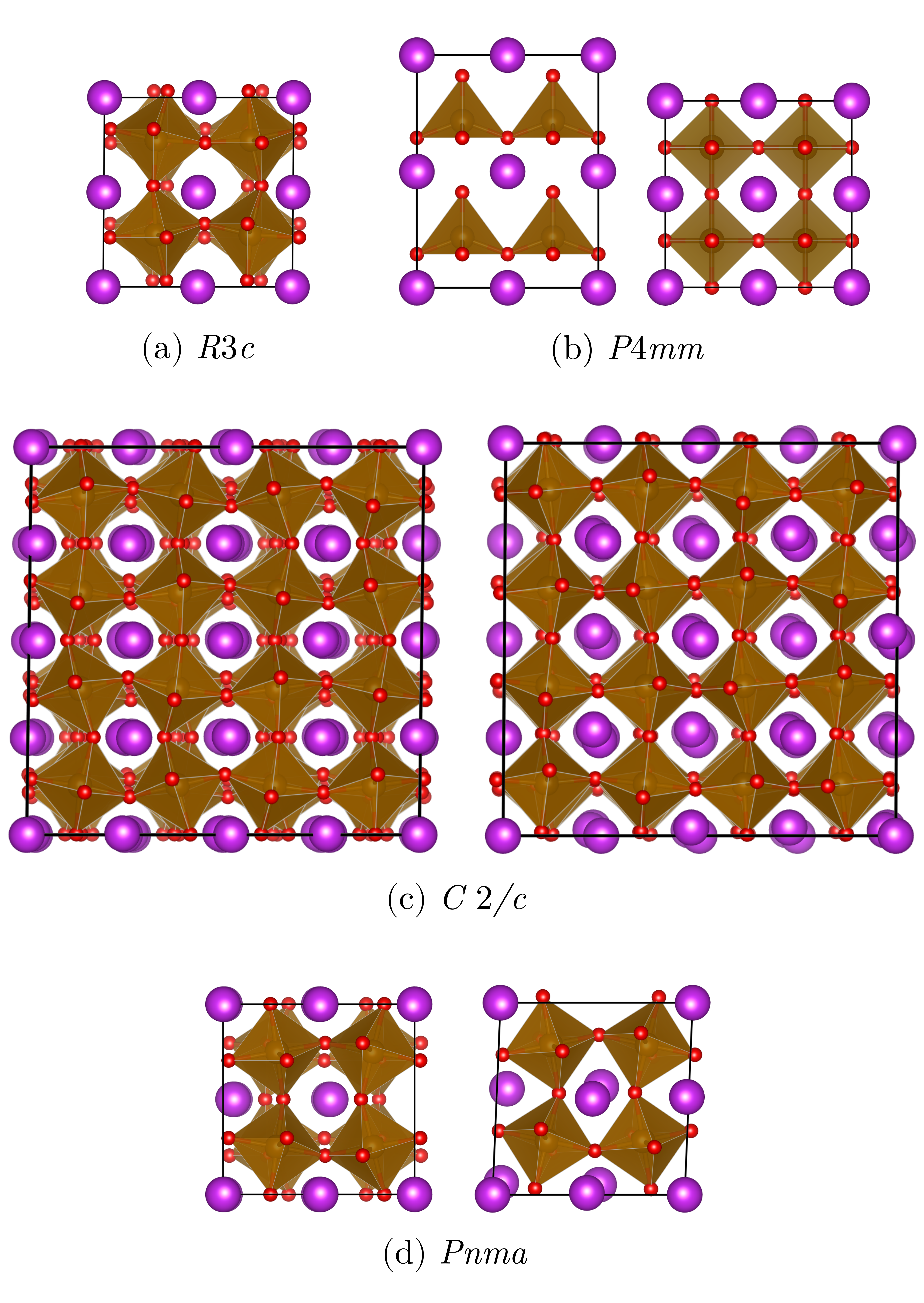} 
\caption{Structures of insulating Bi$X$O$_3$ crystals reported 
experimentally, labelled according to their corresponding space groups:
(a) BiFeO$_3$ at normal conditions,
(b) BiCoO$_3$ at normal conditions,
(c) BiScO$_3$, BiCrO$_3$, and BiMnO$_3$ at normal conditions, and
(d) BiFeO$_3$, BiScO$_3$, BiCrO$_3$, BiMnO$_3$, and BiCoO$_3$ at high
temperature or high pressure.
We plot the smallest pseudocubic supercell compatible with each phase;
here and in later figures
of this kind we show views along the $x$, $y$, and $z$ pseudocubic axes that
are inequivalent (when only two pictures are included, the left picture
corresponds to the two views that are equivalent).}
\label{fig_groundstates}
\end{figure}

BiFeO$_3$ has a simple
crystal structure in bulk at room temperature---a perovskite oxide
with 10 atoms in a rhombohedral unit cell and space group
$R3c$.\cite{Michel1969SSC}
The cations are displaced along the space diagonal of the perovskite
pseudocubic unit cell, and the O$_6$ octahedra rotate in
antiphase about this same three-fold axis, as shown in
Fig.~\ref{fig_groundstates}(a);
the pseudocubic lattice constant is 3.965~\AA~and the pseudocubic angle is
89.45$^\circ$.\cite{Kubel1990AC}
The large displacements of the cations give rise to a
polarization\cite{Catalan2009AM} of around
100~$\mu$C/cm$^2$, while a slight canting of the spins of
the Fe$^{+3}$ ions is responsible for the tiny ferromagnetic moment
experimentally measured in an otherwise strong antiferromagnet of the G
type\cite{Catalan2009AM}
(the type in which
the spins of two nearest Fe$^{+3}$ ions are antiparallel;
in bulk BiFeO$_3$ there is in addition a spiral spin wave of period
640~\AA).

None of the other bismuth $3d$ transition-metal perovskite insulators
displays this $R3c$ structure in bulk.\cite{Belik2012JSSC,Guennou2015CRP}
BiCoO$_3$ 
crystallizes in what is called a supertetragonal structure---a tetragonal
structure of $P4mm$ space group where the $c/a$ ratio is so large (1.27)
that the O$_6$ octahedra become O$_5$ pyramids, as depicted in 
Fig.~\ref{fig_groundstates}(b); every plane of Co$^{3+}$
ions perpendicular to the $c$ axis
has its spin aligned antiparallel to that of its nearest neighbors, while
these planes are stacked along $c$ in ferromagnetic fashion (BiCoO$_3$ is
therefore a C-type antiferromagnet).
BiScO$_3$ (a non-magnetic material), 
BiCrO$_3$ (a G-type antiferromagnet, like BiFeO$_3$), 
and BiMnO$_3$ 
(the only ferromagnet in this family) have been experimentally reported to
crystallize in a centrosymmetric phase with $C2/c$ space group, like the one 
shown in Fig.~\ref{fig_groundstates}(c).
In contrast with this rich behavior under normal conditions,
when the temperature or the pressure becomes large enough all these oxides
(BiScO$_3$, BiCrO$_3$, BiMnO$_3$, BiFeO$_3$, and BiCoO$_3$)
display the same structural phase:\cite{Belik2012JSSC}
the paraelectric structure with space group $Pnma$ commonly found in 
perovksites,\cite{Lufaso2001ACB,Chen2018PRB}
represented in Fig.~\ref{fig_groundstates}(d).

First-principles studies have helped to characterize these structures
for each material involved.
In particular, they have confirmed that they
correspond to special points of the energy surface:
the $R3c$ structure of BiFeO$_3$ was identified as
its ground state according to first-principles calculations already in 
Ref.~\onlinecite{Wang2003S}; McLeod {\em et al.}\cite{McLeod2010PRB} carried
out first-principles optimizations of the $C2/c$ structure 
of BiScO$_3$ and BiCrO$_3$, which showed excellent agreement with experimental
results regarding structural parameters;
Baettig, Seshadri, and Spaldin\cite{Baettig2007JACS} confirmed the $C2/c$ 
structure as the ground state of BiMnO$_3$;
and the supertetragonal structure of bulk BiCoO$_3$ was first analyzed from
the first-principles point of view in Ref.~\onlinecite{Uratani2005JJAP}, where
a large value of its polarization was predicted.
Computational studies also probed phases other than the ground state;
early examples of these concern 
BiCrO$_3$\cite{Baettig2005PRB} (showing that it is possible to optimize
a phase with the structure of bulk BiFeO$_3$)
and BiFeO$_3$\cite{Ravindran2006PRB}
(where the $R3c$ and four other additional phases were optimized).
In our previous work, we reported that the energy surface of bulk BiFeO$_3$
shows many local minima, including
several of supertetragonal type.\cite{Dieguez2011PRB}
We also showed that in BiMnO$_3$ there are
local minima of the bulk energy in addition to the 
experimental $C2/c$ ground state,\cite{Dieguez2015PRB} and that in BiCoO$_3$
the orthorhombic $Pnma$ phase and a $R3c$-like phase have low
energy.\cite{Dieguez2011PRL}
Finally, let us note that there is a considerable amount of work on solid
solutions, particularly lanthanide-doped BiFeO$_{3}$.\cite{Arnold2015TUFFC}
Beyond providing alternatives to tune the properties of bismuth ferrite,
e.g. by inducing morphotropic phases
boundaries\cite{Karpinsky2013JAC,Shi2016JAP,Gonzalez2012PRB} these solid
solutions often feature long-period structures involving unusual rotation
patterns of the O$_{6}$ octahedra.\cite{Prosandeev2013AFM}
 
It is thus abundantly clear that Bi-based perovskite oxides display a
polymorphism that renders them a unique and very attractive materials family. 
In view of this, it is natural to wonder whether these materials have already 
revealed to us all the structures they have in store, or whether they may 
present even more phases with unexpected features. At the same time, it is not 
yet clear whether the polymorphism affects all Bi-based perovskite oxides, 
whether the presence of specific transition metal cations may play a critical 
role, whether a similar structural richness is present (hiding) in other 
perovskites, etc. In this article we address these issues by running a 
systematic search for 
bulk metastable phases of 
Bi$X$O$_3$ compounds ($X={\rm Sc}, {\rm Cr}, {\rm Mn}, {\rm Fe}, {\rm Co}$).
We do this in an automatic and unbiased
way with the help of the {\sc uspex}\cite{uspex1,uspex2,uspex3}
evolutionary algorithm for the search of crystal structures.
We show that these oxides show very similar features in their energy surfaces,
with the same structures appearing as minima in many of the materials, and that
small energy differences between those minima are responsible for
the rich variety of polymorphs displayed.
We also compare this situation with what we obtain for some of the most used
ferroelectric 
perovskite oxides (BaTiO$_3$ and PbTiO$_3$) where very few minima exist 
in their energy surface.
In Section II we describe in detail the
Methodology that we have used. In Section III we present and discuss
our results. Finally, we state our conclusions in Section IV.


\section{Method}
\label{sec:methods}

We used the {\sc vasp} software\cite{vasp} to carry out calculations based
on density-functional theory (DFT).\cite{Hohenberg1964PR,Kohn1965PR}
Since DFT does not reproduce well the localization of
$3d$ electrons in some of these oxides, a correction inspired in 
the Hubbard model was added; this methodology is usually refered to as
LDA+$U$ or DFT+$U$,\cite{Himmetoglu2014IJQC} (details about it are given
in Table \ref{tab_dftu}).
We approximated the exchange-correlation functional following the work 
of Perdew, Burke, and Ernzerhof adapted to solids (PBEsol)\cite{};
different exchange-correlation approximations
can affect the energy difference between minima of the energy 
surface, but in most cases they agree regarding
whether a crystal structure is a minimum of the energy or 
not.\cite{Dieguez2011PRB}
To capture the interaction between the valence electrons and the ion cores
we used projector-augmented wave (PAW) potentials.~\cite{paw1,paw2}
The electrons treated as valence ones were:
3{\it p}, 3{\it d}, and 4{\it s} (Sc, Ti, Cr, Mn, Fe, and Co);
5$s$, 5$p$, 6$s$ (Ba);
5$d$, 6$s$, and 6$p$ (Pb);
5{\it d} and 6{\it s} (Bi);
and 2{\it s} and 2{\it p} (O).

\begin{table}
\caption{
Details of the DFT+$U$ methodology used to treat the $d$ electrons of the
transition-metal atoms: formalism references, values of the 
effective on-site Coulomb interactions $U$, and values of the 
effective on-site exchange interactions $J$.
Those values are chosen based on previous experience;
our tests showed that varying these parameters within a few
eV has a small effect on 
whether a configuration is a minimum of the energy or not, and on the 
variation of the value of
bond lengths and lattice parameters.\cite{Dieguez2017PRB}
Regular DFT (no $U$) was used to treat BiScO$_3$, BaTiO$_3$, and PbTiO$_3$.
}
\begin{tabular}{rcccc}
\hline
\hline
 & BiCrO$_3$ & BiMnO$_3$ & BiFeO$_3$ & BiCoO$_3$
\\
\hline
Formalism
 & Ref.~\onlinecite{Dudarev1998PRB}                 
 & Ref.~\onlinecite{Liechtenstein1995PRB} 
 & Ref.~\onlinecite{Dudarev1998PRB}                 
 & Ref.~\onlinecite{Dudarev1998PRB} 
\\
$U$ & 2.2~eV & 4.0~eV & 4.0~eV & 6.0~eV \\
$J$ & 0.0~eV & 1.0~eV & 0.0~eV & 0.0~eV \\
\hline
\hline
\end{tabular}
\label{tab_dftu}
\end{table}

In order to carry out an unbiased search for energy minima we have used the
{\sc uspex}~\cite{uspex1,uspex2,uspex3} code, which implements an evolutionary
algorithm to find crystal structures with low energy.
We worked with structures that have 10 or 20
atoms in their unit cell; the  initial space group and atomic positions were
chosen randomly by the code.
For every {\sc uspex} run we generated 50 of such random structures, and each
of them was
optimized by {\sc vasp}
using first a conjugate-gradient algorithm and then a quasi-Newton
algorithm.\cite{Pulay1980CPL}
The energy of the resulting configurations
was taken as a fitness parameter to qualify
for the next generation of trial structures, composed of 30 of them 
(this was also the number of
structures used in subsequent generations).
From then on, the population in every generation was computed following
{\sc uspex}'
specially designed variational operators (heredity, mutation, and
soft-mutation).\cite{uspex3}
This procedure was stopped at a maximum of 35 generations, or when the energy
of the best structure did not change for 15 consecutive generations.
The calculations were performed at different degrees of numerical convergence,
with the most accurate level of calculations using a plane-wave energy cutoff
of 500 eV, a reciprocal space resolution of $0.06 \times 2\pi$~\AA$^{-1}$,
and a force stopping criterion of 0.01 eV/\AA.
All the structures obtained in this way were reoptimized using an energy
cutoff of 600 eV, a self-consistent cycle energy threshold of $10^{-8}$ eV,
and a force stopping criterion of 0.001 eV/\AA. 
The zone centered ($\Gamma$-point) frequencies were also calculated using 
finite differences, to ensure that the obtained structures are local minima of
the energy in the simulation box used.

Once a structure was identified by the {\sc uspex} search as a local minimum
of the
energy for one of the materials, we optimized it also for the other materials
(starting with a configuration where the lattice parameters were rescaled
to take into account that the ion sizes are differents for different oxides).
While in some cases {\sc uspex} found the same structures in different
materials, in a few cases these new optimizations identified new energy minima.

As mentioned earlier, the Bi-based oxides studied here show different
magnetic orderings: ferromagnetism (BiMnO$_3$), G-type antiferromagnetism 
(BiCrO$_3$ and BiFeO$_3$), C-type antiferromagnetism (BiCoO$_3$), and
zero magnetic moments (BiScO$_3$).
In our study with {\sc uspex} we used antiferromagnetic 10- and
20-atom unit
cells where the initial spins were assigned randomly.
After we found crystal structures that were a minimum of the energy, we enlarged
(if needed) 
and reoptimized the cell, so that the energy of other typical magnetic orderings
was also computed.
In our previous studies in some these perovskites we
found that once one type of magnetic
ordering is a minimum of the energy, others also exist as
minima of the energy.\cite{Dieguez2011PRB,Dieguez2011PRL,Dieguez2015PRB}

Finally, we resorted to hybrid calculations for the few cases where 
we required a more accurate calculation of the energy differences within a
group of phases, or a more accurate estimation of bandgaps.
We used the modified Heyd-Scuseria-Ernzerhof 
approach\cite{Krakau2006JCP} (HSE06), which predicts bandgaps in good
agreement with experiment for perovskite oxides\cite{Stroppa2010PCCP};
contrary to DFT+$U$, it also predicts the ground state of BiMnO$_3$ to be the
experimentally identified one.\cite{Dieguez2015PRB}
Calculations with this method are two orders of magnitude slower than with
DFT+$U$, but, apart from the better agreement with experiment, they do
not require differente fitting parameters for each material
(HSE06 uses a fraction of exact
exchange equal to 25\%, and a range-separation parameter equal to
0.20~\AA$^{-1}$).

Note that, because we choose to work with relatively small cells, this
investigation does not address the long-period polymorphs that, as mentioned 
above, are believed to occur in some BiFeO$_{3}$-based solid solutions. Yet, 
as we will see, even if restricted to relatively small cell sizes, our 
simulations reveal an incredible structural richness that clearly single out 
Bi-based perovskites from the rest.


\section{Results}
\label{sec:results}

First, we report in detail the BiFeO$_3$ polymorphs identified during our
evolutionary-algorithm search.
Despite of the extensive previous studies carried out for this oxide,
we have identified several structures with low energy that had not been
reported so far as energy minima of BiFeO$_3$.
We then move on to the results of a similar search for 
other Bi-based transition-metal oxides; these harbor similar polymorphs
to those of BiFeO$_3$, but slight differences in energy between those
polymorphs lead to global minima that are structurally very 
different.
We also report similar searches for prototype perovskites
BaTiO$_3$ and PbTiO$_3$ for comparison.
In the following three subsections we give more details about novel
results of our search: structures with energies very close to that
of the ground state of BiScO$_3$, BiCrO$_3$,
and BiMnO$_3$; charge-ordering phases of BiMnO$_3$; and details about a 
polar phase expected to be stable as an epitaxial film is some cases.
Finally, we report bandgap trends for the different phases found, 
and we discuss the reasons for the variety of polymorphs predicted for 
Bi-based perovskite oxides.

\subsection{Local minima of BiFeO$_3$}

Our searches for polymorphs of BiFeO$_3$ with {\sc uspex} following the
procudure describe in the Methods Section produced hundreds of
structures, a result that is consistent with our initial
exploration based on a more rudimentary approach.\cite{Dieguez2011PRB}
In order to sort out which of those structures were unique,
we used the crystal fingerprint method described in
Ref.~\onlinecite{Oganov2009JCP}; this allowed us to identify duplicated
structures that differ slightly in their atomic 
coordinate positions due to numerical noise, and structures that differ only
because of the use of different unit cell shapes and sizes by {\sc uspex}.
The structures that remained were re-optimized in the G- and C-type
antiferromagnetic patterns.
Only the lowest-energy magnetic ordering is reported here.

In this way we found 17 structurally
unique structures within less than 200 meV/f.u.~of the ground
state.
These local minima within 10- and 20-atom unit cells
are listed in Table I.
The ground state is, in agreement with experiment, the polar $R3c$ phase
with G-type antiferromagnetism.
The next configuration in energy is the non-polar $Pnma$ phase; as mentioned
earlier, this is a phase that appears at high
temperature or pressure for Bi-based transition-metal perovskite oxides.

\begin{table*}[ht]
\caption{
BiFeO$_3$ phases found by {\sc uspex} (local minima of the energy in 
10-atom or 20-atom cells).
Column 1: phase index.
Column 2: space group.
Column 3: favoured type of antiferromagnetism.
Column 4: energy difference with the ground state (in meV per formula unit).
Columns 5-10: lattice constants (in~\AA) and lattice angles (in degrees).
Column 11: polarization arising from polar Bi displacements within the
$\Gamma_4^-$ soft-mode of the simple cubic cell (larger
polarization components are denoted as $P$; smaller as $p$).
Column 12: antipolar modes (point of the first Brillouin
zone and direction of displacement of half the Bi cations; the other half move
in the opposite direction).
Column 13: O$_6$ octahedra rotation according to Glazer's 
notation.\cite{GlazerPapers}
We used the {\sc Isotropy} software suite\cite{isotropy} 
to obtain the distortions listed here.
}
\begin{tabular}{ccccllllllcccc}
\hline
\hline
\# & Group & AFM & $\Delta E$
& \multicolumn{6}{c}{Pseudocubic lattice parameters}
& \multicolumn{3}{c}{Significant distortions} \\ 
& & & & $a$ & $b$ & $c$ & $\alpha$ & $\beta$ & $\gamma$ 
& Bi (polar) & Bi (antipolar) & O (rotation) \\ 
\hline
1 & $R3c$ & G & 0
& 3.945 & 3.945 & 3.945 & 89.58 & 89.58 & 89.58 
& $(P_x,P_x,P_x)$
& --
& ($a^-a^-a^-$)
\\
2 & $Pnma$ & G & 28 
& 3.914 & 3.914 & 3.885 & 90    & 90    & 87.64 
& --
& $(2\pi/a)(0,0,1/2)$, $[aa0]$
& ($a^-a^-c^+$)
\\
3 & $C2/c$ & G & 68
& 3.950 & 3.950 & 3.902 & 89.31 & 89.31 & 89.60
& --
& $(2\pi/a)(1/4,1/4,1/4)$, $[aa0]$
& ($a^-a^-c^0$)
\\
4 & $P\bar{1}$ & G & 93
& 3.924 & 3.897 & 3.957 & 89.56 & 89.33 & 89.94
& --
& $(2\pi/a)(1/4,1/4,1/4)$, $[abc]$
& ($a^-b^-c^-$)
\\
5 & $Cc$ & C & 97
& 3.760 & 3.760 & 4.721 & 88.02 & 88.02 & 89.99
& $(p_x,p_x,P_z)$
& $(2\pi/a)(1/2,1/2,1/2)$, $[a\bar{a}0]$
& ($a^-a^-c^0$)
\\
6 & $Pna2_1$ & C & 100
& 3.756 & 3.756 & 4.722 & 90    & 90    & 89.98
& $(0,0,P_z)$
& $(2\pi/a)(0,0,1/2)$, $[aa0]$
& ($a^-a^-c^0$)
\\
7 & $Cm$ & C & 103
& 3.689 & 3.803 & 4.769 & 86.59 & 90    & 90
& $(0,p_y,P_z)$
& $(2\pi/a)(1/2,0,1/2)$, $[0a0]$
& ($a^0b^+c^0$)
\\
8 & $Cc$ & C & 105
& 3.747 & 3.753 & 4.741 & 90    & 87.97 & 90    
& $(p_x,0,P_z)$
& $(2\pi/a)(0,0,1/2)$, $[0a0]$
& ($a^0b^-c^+$)
\\
9 & $Pc$ & C & 106
& 3.750 & 3.750 & 4.744 & 88.09 & 88.09 & 89.74
& $(p_x,p_x,P_z)$
& $(2\pi/a)(1/2,1/2,0)$, $[a\bar{a}0]$
& ($a^0a^0c^+$)
\\
10 & $Pmn2_1$ & C & 106
& 3.690 & 3.799 & 4.762 & 90    & 90    & 90   
& $(0,0,P_z)$
& $(2\pi/a)(0,0,1/2)$, $[0a0]$
& ($a^0b^+c^0$)
\\
11 & $Cm$ & C & 109
& 3.742 & 3.742 & 4.760 & 87.97 & 87.97 & 89.68
& $(p_x,p_y,P_z)$
& --
& --
\\
12 & $Pc$ & G & 109
& 3.987 & 3.987 & 3.971 & 89.06 & 89.06 & 89.92 
& $(P_x,P_x,P_z)$
& $(2\pi/a)(1/2,1/2,0)$, $[a\bar{a}0]$
& ($a^0a^0c^+$)
\\
13 & $Cm$ & G & 111
& 3.915 & 3.928 & 3.982 & 90    & 89.38 & 90   
& $(P_x,0,P_z)$
& $(2\pi/a)(0,1/2,1/2)$, $[a0c]$
& ($a^+b^+c^+$)
\\
14 & $Fmm2$ & C & 116
& 3.747 & 3.753 & 4.675 & 90    & 90    & 90   
& $(0,0,P_z)$
& $(2\pi/a)(1/2,1/2,1/2)$, $[a00]$
& ($a^-b^0c^0$)
\\
15 & $Pmc2_1$ & G & 119
& 4.189 & 4.189 & 3.706 & 90    & 90    & 88.65
& $(P_x,P_x,0)$
& $(2\pi/a)(1/2,1/2,0)$, $[a\bar{a}0]$
& ($a^0a^0c^+$)
\\
16 & $Cm$ & G & 125
& 3.987 & 4.031 & 3.871 & 90    & 90    & 89.18
& $(P_x,P_y, 0)$
& $(2\pi/a)(0,1/2,1/2)$, $[a00]$
& ($a^+b^0c^0$)
\\
17 & $R\bar{3}$ & G & 164
& 3.934 & 3.934 & 3.934 & 89.57 & 89.57 & 89.57
& --
& $(2\pi/a)(1/2,1/2,1/2)$, $[aaa]$
& ($a^-a^-a^-$)
\\
\hline
\hline
\label{tab_bifeo3}
\end{tabular}
\end{table*}

\begin{figure}
\includegraphics[width=70mm]{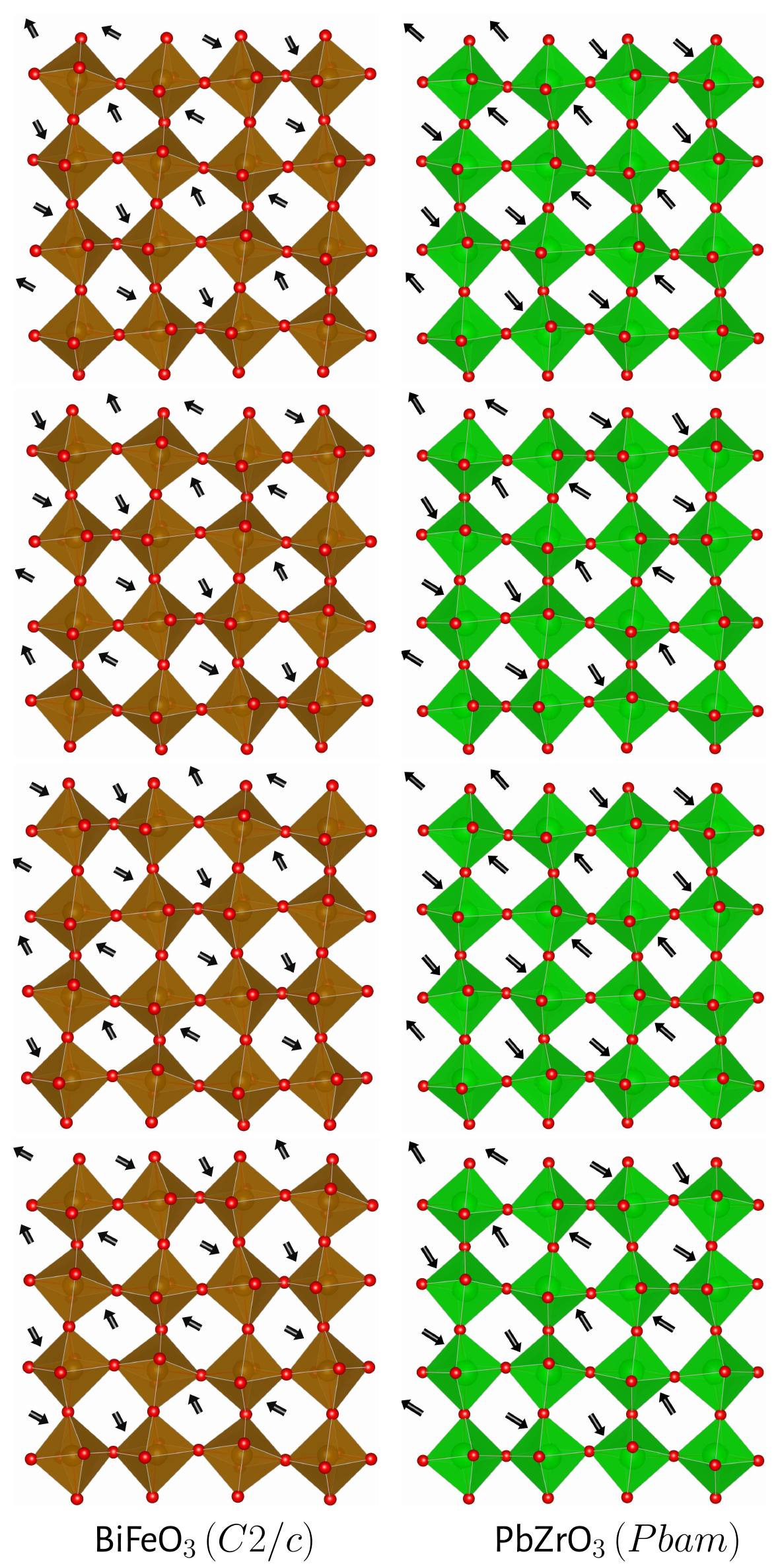} 
\caption{Comparison between the $C2/c$ structure that is a local energy minimum
of 20-atom cell BiFeO$_3$ (left) and the $Pbam$ structure that is a global
energy 
minimum of bulk PbZrO$_3$ (right). Each image shows the distribution of
O$_6$ octahedra in each plane perpendicular to the inequivalent axis of 
each structure, together with arrows that represent the displacement of Bi
atoms in the nearest parallel plane to those octahedra.
}
\label{fig_afe}
\end{figure}

A first surprising result appears next in Table \ref{tab_bifeo3},
as we identified a polymorph of BiFeO$_3$ with relatively low energy that
corresponds to the experimentally reported crystal structure
of bulk BiScO$_3$, BiCrO$_3$, and BiMnO$_3$ in normal conditions.
This is a structure with $C2/c$ space group in which the pattern of 
antiparallel Bi displacements in the planes perpendicular to the inequivalent
axis resemble that of Pb atoms in PbZrO$_3$ (space group $Pbam$),
the prototype of antiferroelectric material.\cite{Trolier2008JAP}
In both structures there is an underlying pattern of O$_6$ rotations in
antiphase along the $[110]$ direction ($a^-a^-c^0$ rotations in Glazer's
notation\cite{GlazerPapers}).
However, as Figure~\ref{fig_afe} shows, the pattern of $A$
cation displacements in the direction perpendicular to the page
is different in the $C2/c$ structure and in the PbZrO$_3$ structure---the
former is
associated with point $(2 \pi/a)(1/4,1/4,1/4)$ of the Brillouin zone of the
simple cubic perovskite cell, while the latter it is associated
with the $(2 \pi/a)(1/4,1/4,0)$ point; for $C2/c$ BiFeO$_3$ the movement
of the Bi atoms distorts further the O$_6$ octahedra.

The similarity with the antiferroelectric phase of PbZrO$_{3}$ suggest an
intriguing possibility: that, if stabilized, the non-polar polymorphs of
BiFeO$_{3}$ would probably behave as antiferroelectrics,\cite{Rabe2013book} 
as it seems likely that they could be transformed, by application of an
electric field, into the polar $R3c$ state. This is in fact a behavior that is
being explored in BiFeO$_{3}$ solid solutions in which the $Pnma$ structure
exists at ambient conditions;\cite{Xu2017NC,Kan2010AFM}
our results suggest that other antiferroelectric phases 
could be similarly identified. (A precise definition of an antiferroelectric 
phase is a matter of some debate; it is not our purpose here to contribute to 
that debate, and we simply adopt the practical view point just introduced. 
Also, in the following, when referring to atomic displacement patterns like 
those of Pb in the PbZrO$_{3}$ structure, we use the terms antiferroelectric 
and antipolar indistinctly.)

When we optimized BiFeO$_3$ starting with a structure like 
that of PbZrO$_3$ we did not obtain a minimum of the
energy---instead, the relaxed structure shows a soft mode that involves
additional antiferroelectric
displacements of the Bi atoms along the inequivalent axis; if we distort
the structure along that mode, the energy indeed lowers, and re-optimization 
leads to a 40-atom unit cell phase around 41 meV/fu above the ground state
with $P2_1/c$ symmetry (which could not have been found in our {\sc uspex}
search since it has a 40-atom primitive cell).

\begin{figure}
\includegraphics[width = 70mm]{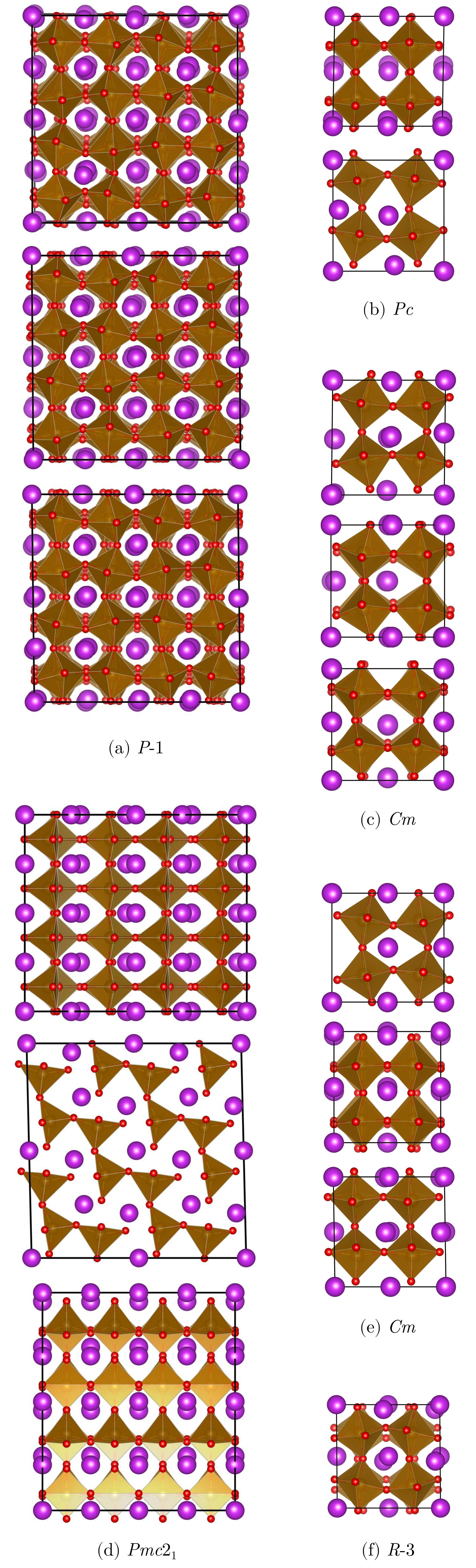} 
\caption{Low energy structures of BiFeO$_3$ displaying G-type 
antiferromagnetism.}
\label{fig_struct2}
 \end{figure}

Complex patterns of antiferroelectric Bi displacements appear also in the
next most stable structure found, of space group $P\bar{1}$ (Figure
\ref{fig_struct2}(a)).
Like the $C2/c$ phase, this low-symmetry non-polar structure shows
an antiferroelectric pattern tied to the $(2 \pi/a)(1/4,1/4,1/4)$ point of the
first Brillouin zone, and
a $a^-b^-c^-$ rotation pattern for the O$_6$ octahedra.
Another difference with the $C2/c$ phase is that now the Bi atoms move
substantially along the three pseudocubic axes.

The four lowest energy structures of Table \ref{tab_bifeo3} have in common
that their three pseudocubic lattice parameters are similar in value.
As discussed in our previous work,\cite{Escorihuela2013PRL} this implies
that the favoured antiferromagnetic ordering is of the G type.
Polar structures 12, 13, 15, and 16, and antipolar structure
17 also fit this pattern, although Fig.~\ref{fig_struct2} shows how diverse
they are in their geometries.

For example, in structure 12 (space group $Pc$, Fig.~\ref{fig_struct2}(b))
we find a mix of a polar
$\Gamma_4^-$ mode
along (111) with and antipolar $M_5^-$ mode along $(1 \bar{1} 0)$, resulting
in large equal Bi displacements along the $z$ pseudocubic axis, and
displacements of Bi atoms in the perpendicular plane that involve half the
atoms moving along the $x$ pseudocubic axis 
and the other half moving along the $y$ pseudocubic axis;
the associated polarization lies close to the (111) pseudocubic axis, while
the O$_6$ octahedra show a $a^0a^0c^+$ pattern.

Structure 13 (space group $Cm$, Fig.~\ref{fig_struct2}(c))
also mixes a polar and an antipolar mode:
a $\Gamma_4^-$ mode along with components along the $x$ and $z$ directions,
and a $M_5^-$ mode almost perpendicular to it; this results in a complex
pattern of
Bi displacements along two different pseudocubic axes and the corresponding
in-plane polarization, together with a $(a^+b^+c^+)$ O$_6$ rotation.

Structure 15 (space group $Pmc2_1$, Fig.~\ref{fig_struct2}(d))
is far from a typical
perovskite; the coexistence of a $\Gamma_4^-$ component along (110) with
a strong $M_5^-$ mode perpendicular to it results in the breaking of the
O$_6$ octahedra into O$_5$ pyramids, but these pyramids form a zig-zag
pattern different to the one in supertetragonal structures.
This structure has been reported before\cite{Yang2012PRL} as a plausible 
phase to appear in epitaxial films under large tensile strains; now we see that
it is a minimum in {\em bulk} BiFeO$_3$, giving support to its
possible experimental realization even at the nominally large strains 
required if comparing to the $R3c$ ground state.
This $Pmc2_1$ phase shows the smallest $c/a$
ratio among the BiFeO$_3$ polymorphs, 
so in the following we refer to it as an {\em inverse supertetragonal}
structure.

Structure 16 (space group $Cm$, Fig.~\ref{fig_struct2}(e))
mixes a polar $\Gamma_4^-$ mode in the $xy$
plane with an antipolar $M_3^-$ mode along $x$, resulting in a net polarization
off the (110) direction.

Finally, structure 17 (space group $R3$, Fig.~\ref{fig_struct2}(e))
is the antipolar equivalent
of the $R3c$ ground state in that for every Bi that moves along 
$(111)$, its six nearest Bi neighbors move along $(\bar1 \bar1 \bar1)$, while
the O$_6$ octahedra pattern is still $(a^-a^-a^-)$.


The rest of phases found by {\sc uspex} are of supertetragonal type.
Structures 5, 6, 7, and 8 where already found in our previous 
search,\cite{Dieguez2011PRB}
while structures 9, 10, 11, and 14 are of a similar kind.
All these supertetragonal structures show a large $\Gamma_4^-$ soft mode
distortion along $z$, and 
some of them show polar Bi displacements in the $xy$ plane 
(5, 7, 8, 9, and 11).
They differ also regarding slight O$_5$ pyramid rotations.

\subsection{Bi$X$O$_3$ Phase Search}
\label{sec:other_compounds}

To help understand the complexity of BiFeO$_3$'s potential energy landscape
we compared it to that of related oxides
Bi$X$O$_3$, with $X = {\rm Sc}, {\rm Cr}, {\rm Mn}, {\rm Co}$.
We run {\sc uspex} on 10-atom and 20-atom unit cells of BiScO$_3$ (spin
unpolarized calculations),
BiCrO$_3$ and BiCoO$_3$ (antiferromagnetic), and
BiMnO$_3$ (both ferromagnetic and antiferromagnetic).
In order to find which magnetic order is favored, we 
re-optimized all minima identified by imposing
ferromagnetic and and A-type antiferromagnetic for BiMnO$_3$,
and G- and C-type for BiCrO$_3$ and BiCoO$_3$ (these are expected to 
be the competitive cases\cite{Dieguez2011PRB,Dieguez2015PRB}).

This search resulted in 22 new minima, to add to the 17 
initial ones of BiFeO$_3$.
The main information about these 39 local minima (in their 10- or 20-atom
unit cell) is contained in
Table~\ref{tab_bixo3}, where their energy is underlined; this energy is 
computed with respect to the latest experimental ground states reported.
Once a structure was found for one of the transition metals, we re-optimized
it for the others too; if this led us to a energy minimum of similar
geometric characteristics, we 
quote its energy, space group, and magnetic ordering in Table~\ref{tab_bixo3}
(not underlined).
For all calculations reported in this Table we extended the unit cell of
the structures to a 40-atom unit cell---the $2 \times 2 \times 2$ pseudocubic
one when possible,
and the conventional cell of the $C2/c$ phase for the other cases.
The existence of negative eigenvalues of the dynamical matrix was checked 
for those extended 40-atom unit cells (and not only for the cell in which
the minima was found, as in Table~\ref{tab_bifeo3}).
If soft modes exist, we give the energy of the structure
in brackets.
Thus, for example, structures 3, 11, and 16 are reported as local minima
in Table~\ref{tab_bifeo3} because they are so in their respective 20-, 10-, and
20-atom unit cells where they were found by {\sc uspex}; however, when
we used 40-atom unit cells soft modes appeared in those three cases.
The structure that results when we follow the soft mode of the $C2/c$ phase
is the subject of next subsection.

\begin{table*}
\caption{
Characteristics of the structures found after our search 
over 10-atom and 
20-atom unit cells of five Bi-based transition-metal perovskite oxides.
For each material there are four columns: S.G. contains the space group
of the structure; M. contains the lowest energy magnetic configuration
(one capital letter indicates antiferromagnetic type, 
FM indicates ferromagnetism---an asterisk is added if there
is charge ordering); $\Delta E$ is the difference in energy
(in meV/f.u.) with the structure that is the accepted experimental ground
state; and $c/a$
is the ratio between the most dissimilar lattice parameter and the average of
the other two (in the perovskite pseudocubic unit cell).
When the energy value is underlined, it means that this structure was 
found by the {\sc uspex} code; otherwise, it was found by adapting a
structure found by {\sc uspex} in one of the other materials.
When the energy value is in brackets, it means that this is not a local minimum
of the energy in the corresponding 40-atom pseudocubic cell, but a saddle 
point.
Structures in the same line are considered to be the same (they show the
same main distortions in terms of cation displacements and O$_6$ rotations),
even if in some cases Jahn-Teller distortions cause the space group to vary.
}
\begin{tabular}{rcccrcccrcccrcccrccc}
\hline
\hline
  \multicolumn{4}{c}{BiScO$_3$}
& \multicolumn{4}{c}{BiCrO$_3$}
& \multicolumn{4}{c}{BiMnO$_3$}
& \multicolumn{4}{c}{BiFeO$_3$}
& \multicolumn{4}{c}{BiCoO$_3$}
\\
  S.G. & M. & $\Delta E$ & $c/a$ 
& S.G. & M. & $\Delta E$ & $c/a$ 
& S.G. & M. & $\Delta E$ & $c/a$ 
& S.G. & M. & $\Delta E$ & $c/a$ 
& S.G. & M. & $\Delta E$ & $c/a$ 
\\
\hline
  $P4mm$     & 0  &           (409) & 1.296
&            &    &                 &   
& $P4mm$     & A  &           (390) & 1.264
& $P4mm$     & C  &           (140) & 1.282
& $P4mm$     & C  & \underline{0}   & 1.257
\\
  $R3c$      & 0  & \underline{-45} & 1.000
& $R3c$      & G  & \underline{-22} & 1.000
& $Cc$       & FM & \underline{7}   & 0.975
& $R3c$      & G  & \underline{0}   & 1.000
& $Cc$       & G  & \underline{38}  & 1.026
\\
  $Pnma$     & 0  & \underline{-35} & 0.994
& $Pnma$     & G  &            -59  & 0.996
& $Pnma$     & FM &            -17  & 0.962
& $Pnma$     & G  & \underline{28}  & 0.992
& $Pnma$     & G  &            47   & 1.014
\\
             &    &                 &      
&            &    &                 &      
& $C2/c$     & FM*& \underline{-15} & 0.985          
&            &    &                 &       
&            &    &                 &       
\\
             &    &                 &   
&            &    &                 &   
& $P2_1/c$   & FM*& \underline{-9}  & 0.990
&            &    &                 &   
&            &    &                 &   
\\
  $C2/c$     & 0  &             0   & 0.984
& $C2/c$     & G  &            (0)  & 0.993
& $C2/c$     & FM &            (0)  & 1.017
& $C2/c$     & G  &(\underline{68}) & 0.988
& $C2/c$     & G  &            53   & 0.979
\\
             &    &                 &      
&            &    &                 &      
& $R3$       & FM*& \underline{2}   & 1.000
&            &    &                 &      
&            &    &                 &      
\\
             &    &                 &      
&            &    &                 &      
& $R\bar{3}$ & FM*& \underline{20}  & 1.000
&            &    &                 &        
&            &    &                 &        
\\
  $Pna2_1$   & 0  & \underline{-25} & 1.009     
&            &    &                 &      
&            &    &                 &      
&            &    &                 &      
&            &    &                 &      
\\
  $P\bar{1}$ & 0  &            28   & 1.011     
& $P\bar{1}$ & G  &           (26)  & 0.994
& $P\bar{1}$ & FM &           (37)  & 1.019
& $P\bar{1}$ & G  & \underline{93}  & 1.012
& $P\bar{1}$ & G  &            88   & 0.964
\\
  $R\bar{3}$ & 0  & \underline{127} & 1.000     
& $R\bar{3}$ & G  &             43  & 1.000
& $P\bar{1}$ & FM &             64  & 0.952
& $R\bar{3}$ & G  & \underline{164} & 1.000
& $P\bar{1}$ & G  & \underline{147} & 1.023     
\\
  $Cm$       & 0  & \underline{63}  & 0.988     
& $Cm$       & G  &           (95)  & 1.003
& $Cm$       & FM &            97   & 0.949
& $Cm$       & G  & \underline{111} & 1.015
& $Cm$       & G  &           (139) & 1.025     
\\
             &    &                 &      
&            &    &                 &      
&            &    &                 &      
& $Cc$       & C  & \underline{97}  & 1.255
&            &    &                 &      
\\
             &    &                 &      
&            &    &                 &      
& $Pna2_1$   & A  &            139  & 1.149
& $Pna2_1$   & C  & \underline{100} & 1.257
&            &    &                 &      
\\
             &    &                 &      
&            &    &                 &      
&            &    &                 &      
& $Cm$       & C  & \underline{103} & 1.273
&            &    &                 &      
\\
             &    &                 &      
&            &    &                 &      
&            &    &                 &      
& $Cc$       & C  & \underline{105} & 1.264
&            &    &                 &      
\\
             &    &                 &      
&            &    &                 &      
& $Pc$       & A  &            212  & 1.220
& $Pc$       & C  & \underline{106} & 1.265
&            &    &                 &      
\\
             &    &                 &      
&            &    &                 &      
&            &    &                 &      
& $Pmn2_1$   & C  & \underline{106} & 1.272
&            &    &                 &      
\\
  $Cc$       & 0  & \underline{61}  & 1.022
&            &    &                 &
&            &    &                 &      
&            &    &                 &        
& $Cc$       & G  &           107   & 1.039
\\
  $Pc$       & 0  & \underline{63}  & 0.991
&            &    &                 &
&            &    &                 &
&            &    &                 &
&            &    &                 &
\\
             &    &                 &      
&            &    &                 &      
&            &    &                 &      
& $Cm$       & C  &(\underline{109})& 1.272
&            &    &                 &      
\\
  $Pc$       & 0  &            69   & 1.004
& $Pc$       & G  &           (122) & 1.015     
& $Pc$       & FM &           (127) & 1.000     
& $Pc$       & G  & \underline{110} & 0.996
& $Pc$       & G  &           (148) & 1.025
\\
  $Pna2_1$   & 0  & \underline{64}  & 0.997     
& $Pna2_1$   & G  & \underline{125} & 1.001
& $Pna2_1$   & FM &            122  & 0.990
& $Pna2_1$   & G  &            126  & 0.992
& $Pna2_1$   & G  &            136  & 1.020
\\
  $Ama2$     & 0  &(\underline{65}) & 0.996
&            &    &                 &
& $Ama2$     & FM &          (131)  & 0.988
&            &    &                 &
&            &    &                 &
\\
  $Pmc2_1$   & 0  &           (208) & 0.879     
&            &    &                 &      
& $Pmc2_1$   & FM &           (146) & 0.907
& $Pmc2_1$   & G  & \underline{119} & 0.885
&            &    &                 &           
\\
  $Pmc2_1$   & 0  &           (236) & 0.970     
& $Pmc2_1$   & G  &           (169) & 1.014
&            &    &                 & 
& $Pmc2_1$   & G  &           (163) & 0.969
& $Pmc2_1$   & G  &(\underline{111})& 1.014     
\\
             &    &                 &      
&            &    &                 &      
&            &    &                 &      
& $Fmm2$     & C  & \underline{116} & 1.247
&            &    &                 &      
\\
  $Cm$       & 0  &            (81) & 0.968
& $Cm$       & G  &           (118) & 1.024
& $Cm$       & FM &           (121) & 1.039
& $Cm$       & G  &(\underline{125})& 0.966
& $Cm$       & G  &           (117) & 1.031
\\
             &    &                 &      
&            &    &                 &
&            &    &                 &      
&            &    &                 &        
& $Cc$       & G  & \underline{144} & 1.041
\\
  $Cm$       & 0  & \underline{102} & 1.009
&            &    &                 &
&            &    &                 &
&            &    &                 &
&            &    &                 &
\\
\hline
\hline
\label{tab_bixo3}
\end{tabular}
\end{table*}

Another example will help to clarify how information is reported in 
Table~\ref{tab_bixo3}.
For BiCoO$_3$, {\sc uspex} found the supertetragonal $P4mm$ phase that is 
indeed the known experimental ground state of the material; so an 
underlined  value of 0 meV/fu is added to Table~\ref{tab_bixo3} in the
BiCoO$_3$ column.
When computing energies of different magnetic arrangements, antiferromagnetism
of the C type is
favored (also in agreement with experiment), so this is too reflected, together
with the large $c/a$ ratio.
We then run the same structure for the other four materials (we started 
optimizations by simply changing the $B$ cation and using a initial cell 
volume adjusted proportionally to the size of the new cation).
In three cases (BiScO$_3$, BiMnO$_3$, and BiFeO$_3$) the optimization 
converged, but there were negative dynamical matrix eigenvalues in the 
$2 \times 2 \times 2$ pseudocubic cell, so the energy values are in brackets. 
In the other case, BiCrO$_3$, the supertetragonal $P4mm$ phase would just not
be a special point of the energy surface.

Once again, we see a richness of polymorphs that is
unique to bismuth perovskite oxides.
The phases here include the previously reported ones for these five 
materials, most notably their ground state form according to experiment and 
the $Pnma$ phase that they all reach at high enough temperature or pressure.
We can establish three groups of structures: phases 
with similar pseudocubic lattice constants, phases with one pseudocubic
lattice constant significantly larger than the other two
(supertetragonal phases), 
and phases with one pseudocubic lattice constant significantly smaller
than the other two
(inverse supertetragonal phases).
Among the first group, a phase that appears for all five materials as a 
local minimum is the $R3c$ phase (global minimum of BiFeO$_3$).
In BiMnO$_3$ and BiCoO$_3$ this phase is slightly distorted because Mn$^{+3}$
and Co$^{+3}$ are Jahn-Teller active ions, so their space group becomes
actually $Cc$; however, the main distortions in the structures are still the
large cation displacement along the (111) pseudocubic direction and a
O$_6$ rotation similar to $a^-a^-a^-$,
so we group these minima in the same line of the Table.
The supertetragonal polar phases appear only in BiMnO$_3$, BiFeO$_3$ and
BiCoO$_3$, with A-antiferromagnetic ordering (BiMnO$_3$) and
C-antiferromagnetic ordering (BiFeO$_3$ and BiCoO$_3$).
On the other hand, non-supertetragonal polar and non-polar phases 
can be seen in all five materials; they favour G-antiferromagnetic ordering in 
BiCrO$_3$, BiFeO$_3$, and BiCoO$_3$, while ferromagnetic ordering and 
A-antiferromagnetic ordering are very close in BiMnO$_3$, as 
previously reported.\cite{Dieguez2015PRB}
The inverse supertetragonal phases are analyzed in more detail in subsection E.

Apart from the high number of structures listed, a few other facts 
deserve to be commented in relation to Table~\ref{tab_bixo3}.
First, there are several instances of negative energies for BiScO$_3$,
BiCrO$_3$, and BiMnO$_3$.
This implies that according to our calculations, those phases have energies
below that of the assumed experimental ground state. 
In a previous article we reported that for BiMnO$_3$ DFT+$U$ does not
agree with experiment in that the $C2/c$ phase is the one with the lowest
energy, but that when using hybrid methods reconciliation with experiment
is achieved. 
Now we see that the situation is extended to BiScO$_3$ and BiCrO$_3$.
In next subsection we look in more depth at this issue.
Second, we have found that BiMnO$_3$ shows some unique phases where 
charge separation is present---Mn$^{+2}$ and Mn$^{+4}$ ions coexist.
This is further discussed in subsection D.

As a point of comparison, we also run analogous {\sc uspex} optimization
calculations for prototypical perovskites BaTiO$_3$ and PbTiO$_3$.
Full runs as those described in the Methods section produced just the 
known ground states of these materials: the rhombohedral $R3m$ phase for
BaTiO$_3$ and the tetragonal $P4mm$ phase for PbTiO$_3$.
No other local minima was found in 20-atom unit cells
(the tetragonal $P4mm$ and orthorhombic $Amm2$ phases of BaTiO$_3$ are
saddle points, not minima).

\begin{figure}
\includegraphics[width=60mm]{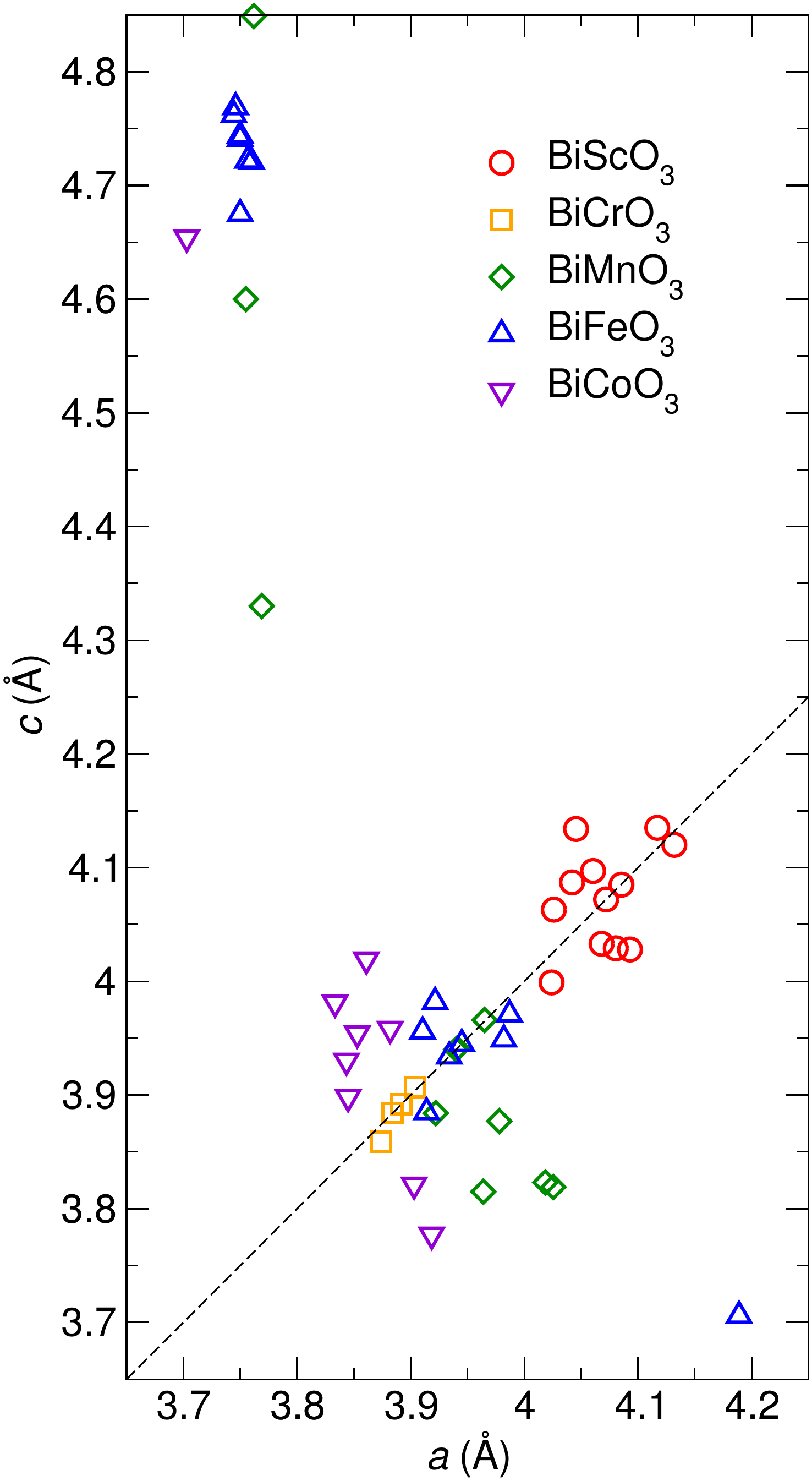} 
\caption{
Out-of-plane lattice parameter $c$ versus in-plane lattice parameter $a$
(the average of the two closest lattice parameters in the pseudocubic
setting),
for the structures listed in Table~\ref{tab_bixo3} that are local minima of
the energy.
The discontinuous line corresponds to $c=a$.
}
\label{fig_tableII_cVSa}
\end{figure}

To give a graphical idea of the richness of Bi$X$O$_3$ phases found,
Fig.~\ref{fig_tableII_cVSa} shows the value of the $c$ lattice parameter versus
the average $a$ and $b$ parameters for those phases of Table \ref{tab_bixo3}
that are local minima. 
The lattice parameters are sorted so that $a$ and $b$ are as close as possible
to each other; we do this with the idea in mind that a possible way to access
local minima of these materials is to grow them as epitaxial 
films on square perovskite lattices---then, the in-plane lattice parameter
would likely be around $(a+b)/2$ and the out-of-plane lattice parameter would
be around $c$.
It is apparent from the graph that one cluster of supertetragonal phases
appear at the top left of the figure, while most of the rest of the phases
have $c/a$ ratios not far from one.
The outlier at the bottom right part of the graph is phase 15 of BiFeO$_3$
(in this graph only phases that are minima in the 40-atom unit cells described
earlier are represented, so the inverse supertetragonal phases of other
materials are absent).

This adaptability of the Bi ion to different environments of perovskite
variations implies not only that phases other than the ground state might
be stabilized by strain or pressure, but also that these materials should
have richer surfaces and interfaces than typical perovskites.
For example, some BiFeO$_3$ domain walls can be seen as narrow regions 
in which the structure corresponds to a diffferent
polymorph\cite{Dieguez2013PRB,Prosandeev2013AFM} 
and the same it is true for its surfaces.\cite{Marti2011PRL}

\subsection{Ground State of BiScO$_3$, BiCrO$_3$, and BiMnO$_3$}

We move on now to the issue of the phases of BiScO$_3$, BiCrO$_3$, and 
BiMnO$_3$ that show lower energies than the one of the experimental ground
state of these materials, the phase with $C2/c$ symmetry.
In our previous study of BiMnO$_3$\cite{Dieguez2015PRB} we reported that 
DFT+$U$ calculations very similar to the ones presented here indeed 
stablish that the $Pnma$ phase has a somewhat lower energy than the $C2/c$
phase.
We also showed that when using hybrid functionals, this ordering is reversed,
and the $C2/c$ phase becomes the lowest-energy one, in accordance to the
latest experiments about the structure of this material.\cite{Belik2012JSSC}
Now we see that for BiScO$_3$ and BiCrO$_3$ we also obtain that the $Pnma$
phase has lower energy than the $C2/c$ phase.
Moreover, in this case this is also true for the $R3c$ phase. 

More puzzingly, the data in Table~\ref{tab_bixo3} tells that the $C2/c$ phase
of BiCrO$_3$ and BiMnO$_3$ shows a soft mode when the conventional 40-atom
unit cell is used.
When we perturb slightly the atomic positions of these structures by following
the eigenvector of the soft mode, indeed the energy of BiCrO$_3$ and BiMnO$_3$ 
(and also BiFeO$_3$) goes down, while this does not happen in BiScO$_3$
(and BiCoO$_3$).
This is consistent with the structure being a minimum in BiScO$_3$ (and
BiCoO$_3$), but a saddle point in BiCrO$_3$ and BiMnO$_3$ (and BiFeO$_3$).
What is the structure that appears if we keep following the eigenvector of
the soft mode and then let the atoms relax?
This structure has 40 atoms in its unit cell (and therefore could not have
been found in our {\sc uspex} search over 10-atom and 20-atom cells), and
space group $P2_1/c$.
In addition to the distortions quoted in Table~\ref{tab_bifeo3} for the 
$C2/c$ phase, this $P2_1/c$ phase 
has two other prominent ones: an antipolar Bi displacement
along the $(001)$ pseudocubic axis, and a $(a^0 a^0 c^+)$ O$_6$ rotation.
In all, it shows a similar O$_6$ rotation pattern to that of the $Pnma$ phase,
and a complicated antipolar Bi pattern associated mainly to the 
$(2\pi/a)(1/4,1/4,1/4)$ point of the simple cubic first Brillouin zone.

\begin{figure}
\includegraphics[width=77mm]{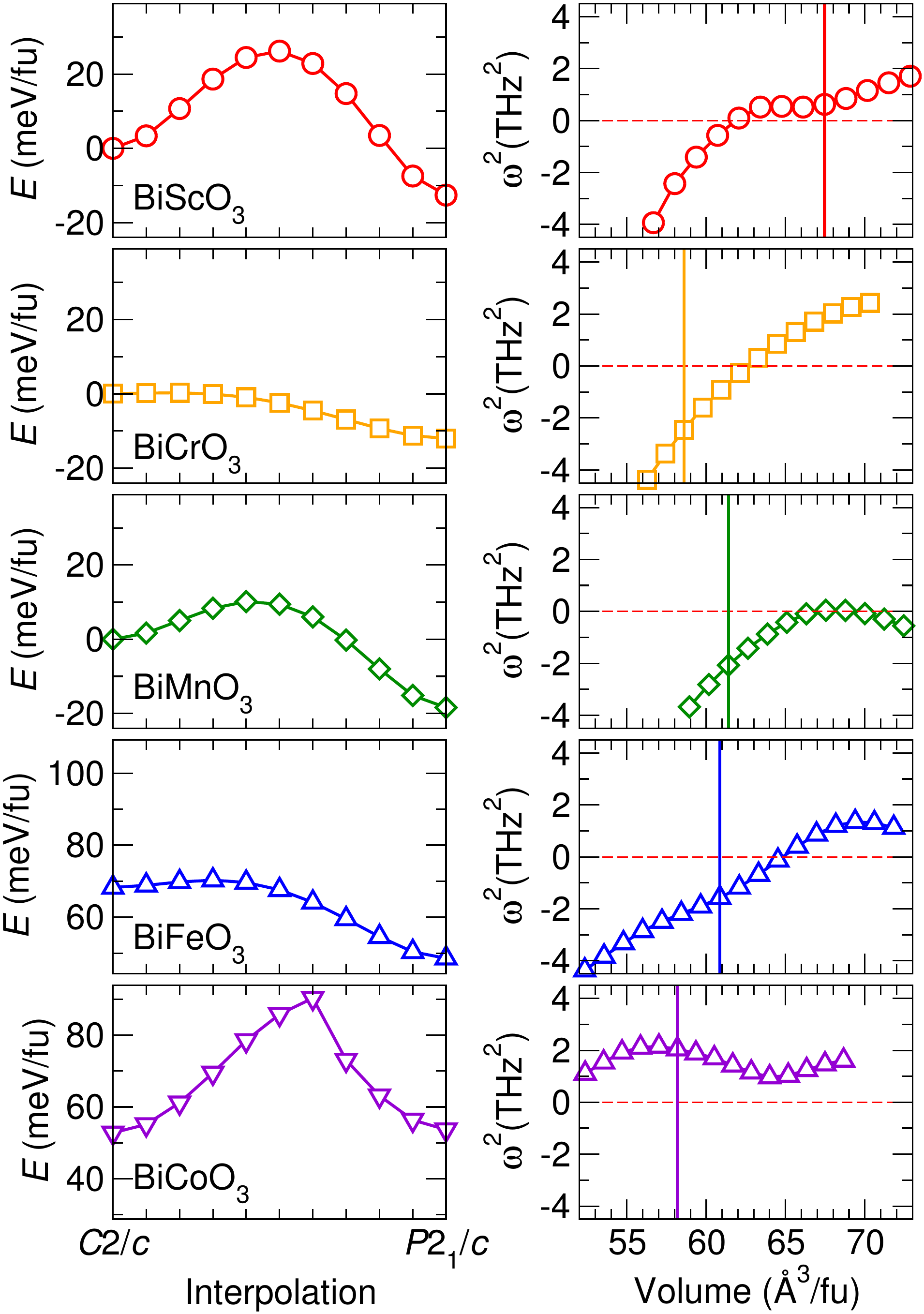}
\caption{
Left panels:
energy of the structures that result from a linear interpolation of atomic
positions and lattice vectors between the $C2/c$ phase and the $P2_1/c$ phases
of the materials quoted.
Right panels:
square of the frequency of the mode that connects the $C2/c$ and $P2_1/c$ 
phases as a function of the volume of the $C2/c$ unit cell; 
the vertical lines mark the volume of the cell with optimal lattice parameters.
}
\label{fig_c2c_to_p21c}
\end{figure}

Figure~\ref{fig_c2c_to_p21c} (left panels) shows the energy as we interpolate
linearly the atomic positions and lattice vectors from the $C2/c$ phase
to the $P2_1/c$ phase, for all five Bi-based perovskites.
It is apparent that the curvature at the $C2/c$ point is much larger for
those materials where this phase is a minimum (BiScO$_3$ and BiCoO$_3$) than
for the others.
When the $C2/c$ phase is a saddle point of the energy surface, then the
$P2_1/c$ phase corresponds to a minimum with lower energy;
when instead the $C2/c$ phase 
is a minimum, then the $P2_1/c$ phase is also a minimum, but it might exist
at a lower energy (BiScO$_3$) or at a higher one (BiCoO$_3$). 

What is the reason for the different character of the $C2/c$ phases in the
different oxides analyzed here?
To help us answer this question, Figure~\ref{fig_c2c_to_p21c} (right panels)
contains the value of the frequency of the mode that becomes soft in 
BiCrO$_3$ and BiMnO$_3$, for different values of unit cell volume---i.e.,
we increased the length of the $a$, $b$, and $c$ lattice parameter of the
$C2/c$ conventional unit cell by some amount, relaxed the atoms with those
fixed lattice vectors, computed the eigenvalues of the dynamical matrix, 
and plotted the square of the eigenvalue of interest in the graphs.
What results is consistent
with a similar behavior of this frequency for all five materials:
at large values of volume there is room for the O$_6$ cages without the 
need for the extra rotation of the $P2_1/c$ phase, so the eigenvalues
are positive, while at small enough volumes this is not the case and the
structure adds those rotations as a way to avoid compressing the bonds between
transition metals and oxygens further.
In between those regimes, the rotations and the Bi displacements compete,
giving rise to the non-monotonic behaviour seen in the panels.
Depending on what the optimal value of the volume is for a given material,
different parts of the curve are accessed, and the different behaviours 
listed earlier result.

The previous discussion points out that the relative ordering of the
$C2/c$ and $P2_1/c$ phases is a subtle effect.
To further investigate this issue, and the one of the negative energies 
of Table~\ref{tab_bixo3}, we have used hybrid functionals.
These computationally demanding calculations represent today one of the most
accurate types of first-principles calculations that can be performed to
compute energies for unit cells of 40
atoms.\cite{Krakau2006JCP,Stroppa2010PCCP}
The results we obtained are quoted in Table~\ref{tab_hybrid}.
Some of the BiMnO$_3$ data had been reported before,\cite{Dieguez2015PRB}
showing that the $C2/c$ phase was lower in energy than the $R3c$ and $Pnma$
phases; 
now we see that the $P2_1/c$ phase has very close energy to the $C2/c$ phase,
but $C2/c$ is still the ground state according to hybrid calculations.
Also in BiScO$_3$ the $C2/c$ phase is the ground state according to our HSE06
calculations, in agreement with experiment.
For BiCrO$_3$, however, the hybrid calculations put the analyzed structures
in a small bracket of energies where the $C2/c$ phase is not the lowest-energy
one (whether this is an artifact or not of the calculations is difficult
to assess, given the small differences in energy involved).
Computing possible phonon instabilities of these phases with the HSE06 hybrid
functional 
would require ten to twenty times more computer power than
the calculations reported in Table~\ref{tab_hybrid}, which are already
computationally demanding.
Instead, the results reported in Fig.~\ref{fig_c2c_to_p21c} give evidence
that when the $C2/c$ phase is lower in energy than the $P2_1/c$ phase both
correspond to energy minima.
When using hybrids we see that the $C2/c$ phase in BiScO$_3$ and BiMnO$_3$
is lower in energy than the $P2_1/c$ phase, so we expect both structures
to be minima ($C2/c$ being the global one, in agreement with
experiment).

\begin{table}
\caption{
Results of HSE06 calculations of the most stable structures of bulk 
BiScO$_3$, BiCrO$_3$, and BiMnO$_3$.
For each material we report its lowest magnetic ordering, the energy 
difference with the $C2/c$ phase (in meV/f.u.), and its band gap (in eV). 
}
\begin{tabular}{rccccccccc}
\hline
\hline
& \multicolumn{3}{c}{BiScO$_3$}
& \multicolumn{3}{c}{BiCrO$_3$}
& \multicolumn{3}{c}{BiMnO$_3$} \\
 Phase & M. & $\Delta E$ & Gap & M. & $\Delta E$ & Gap & M. & $\Delta E$ & Gap
\\
\hline
        $C2/c$ &   0 &   0 & 4.3 &   G &   0 & 3.1 &  FM &   0 & 1.8 \\
      $P2_1/c$ &   0 &  14 & 4.0 &   G &  -2 & 3.1 &  FM &   1 & 1.9 \\
 $R3c$ or $Cc$ &   0 &   4 & 4.2 &   G & -13 & 3.3 &   A &  33 & 2.5 \\
        $Pnma$ &   0 &  64 & 4.0 &   G &  -8 & 2.7 &  FM &  53 & 1.7 \\
\hline
\hline
\label{tab_hybrid}
\end{tabular}
\end{table}

\subsection{Charge-Ordering Phases in BiMnO$_3$}

\begin{figure}
\includegraphics[width=40mm]{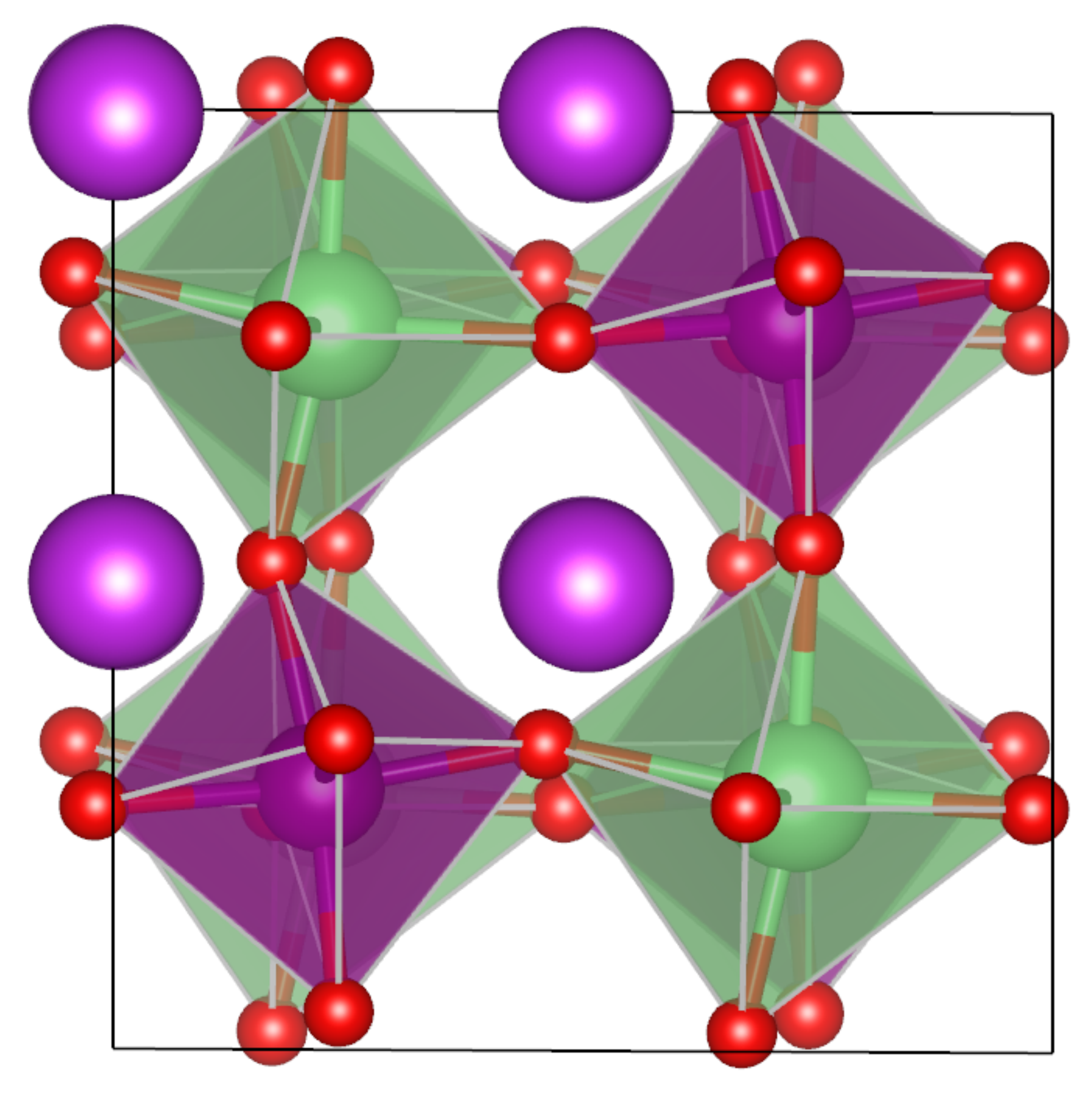}
\includegraphics[width=40mm]{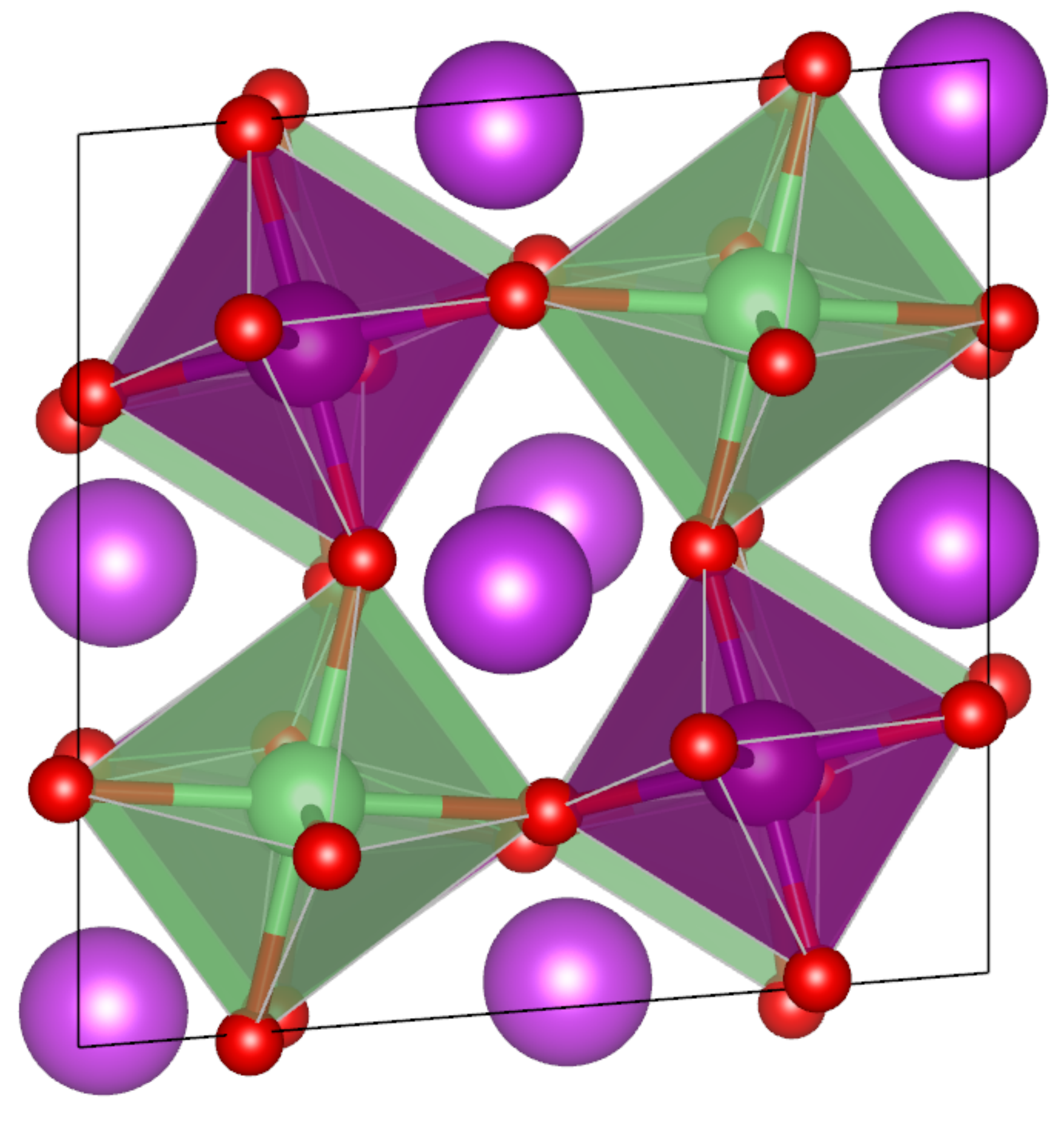}
\caption{
40-atom unit cells for two of the charge-ordering phases of BiMnO$_3$ found
in this work: a $R3c$-like phase (left) similar to the ground state of 
BiFeO$_3$ but with a rock-salt pattern imposed on the perovksite $B$ sites,
and a $Pnma$-like phase (right) similar to the high-pressure phase of these
materials but with the same rock-salt pattern.
The larger (smaller) octahedra enclose Mn$^{+2}$ (Mn$^{+4}$) ions.
}
\label{fig_co}
\end{figure}

Another interesting feature reported in Table~\ref{tab_bixo3} is the presence
of four phases of BiMnO$_3$ that possess charge ordering.
These correspond, in order of increasing energy, to variations of the
$C2/c$, $Pnma$, $R3c$, and $R\bar{3}$ phases, where a rocksalt pattern 
of Mn$^{2+}$ and Mn$^{4+}$ ions exists in the $B$ site of the perovksite
(unit cells of two of the phases found are pictured in Fig.~\ref{fig_co}).
Charge-ordering phases of perovskites have been reported 
experimentally before, with the ordering taking place either in the
$B$ site (e.g., Mn$^{+3}$ and Mn$^{+4}$ in
Ln$_{1-x}$$A_x$MnO$_3$, where Ln $=$ rare earth and $A =$ Ca, 
Sr\cite{Rao1998CM}) or in the $A$ site (e.g., Bi$^{+3}$ and Bi$^{5}$ in 
BiNiO$_3$\cite{Ishiwata2002JMC}).
However,
no $AB$O$_3$ perovskite with charge ordering in $B$ seems to have been found
so far.

To check how robust the prediction done with DFT+$U$ is, we
also carried out hybrid calculations for these phases (assuming that the
spins are oriented parallel to each other).
Table~\ref{tab_co} contains the results of a comparison of the structure of 
these variations and the one of the regular phases.
In particular, we show the average Mn--O distance obtained for each individual
Mn ion of the structure; these results agree well with the picture of ionic
bonding that emerges from the
radii given by Shannon\cite{Shannon1976AC}
(1.35~\AA~for O$^{-2}$,
0.53~\AA~for Mn$^{+4}$,
0.65~\AA~for Mn$^{+3}$,
0.82~\AA~for Mn$^{+2}$).

\begin{table}
\caption{
Results of HSE06 calculations for some of the regular phases of BiMnO$_3$,
and for its charge-ordering phases.
For each material we report the energy 
difference of the given phase (parallel spins) with respect to the $C2/c$ phase
(in meV/f.u.), the lattice parameters, and the average Mn--O distances.
}
\begin{tabular}{rccccccc}
\hline
\hline
&
& \multicolumn{3}{c}{Latt. param.}
& \multicolumn{3}{c}{Mn--O dist. (\AA)} \\
 Phase & $\Delta E$ & $a$~(\AA) & $b$~(\AA) & $c$~(\AA)
       & Mn$^{4+}$ & Mn$^{3+}$ & Mn$^{2+}$
\\
\hline
\multicolumn{8}{l}{\em Regular phases:} \\
          $C2/c$ &   0 & 3.903 & 3.903 & 3.975 &  --  & 2.02 &  --  \\
      $R3c$-like &  36 & 4.026 & 4.026 & 3.772 &  --  & 2.02 &  --  \\
          $Pnma$ &  53 & 3.949 & 3.949 & 3.764 &  --  & 2.03 &  --  \\
 $R\bar{3}$-like & 102 & 3.986 & 4.072 & 3.749 &  --  & 2.02 &  --  \\
\multicolumn{8}{l}{\em Phases with charge ordering:} \\
     $C2/c$-like &  97 & 3.929 & 3.929 & 3.857 & 1.91 &  --  & 2.13 \\
      $R3c$-like & 152 & 3.919 & 3.919 & 3.919 & 1.92 &  --  & 2.09 \\
     $Pnma$-like & 160 & 3.923 & 3.923 & 3.870 & 1.92 &  --  & 2.12 \\
 $R\bar{3}$-like & 145 & 3.918 & 3.918 & 3.918 & 1.91 &  --  & 2.13 \\
\hline
\hline
\label{tab_co}
\end{tabular}
\end{table}

As seen in Table~\ref{tab_co}, the energy differences between the phases
with charge ordering and the ground state are larger than the ones reported
in Table~\ref{tab_bixo3} using DFT+$U$.
Still, these differences are similar to the ones between
experimentally grown supertetragonal
phases of BiFeO$_3$ and its ground state.
It might therefore be feasible to stabilize
some of these phases;
a possible route could be to grown BiMnO$_3$ epitaxially on a
perovskite substrate that shows a checkerboard pattern compatible with the
distortions, e.g., using as substrate one of the many
ordered double perovskites that are known.\cite{King2010JMC}

\subsection{Highly Strained Epitaxial Film Phase}

In Fig.~\ref{fig_tableII_cVSa} we plotted the behaviour of the lattice 
parameter that most differ from the other two ($c$) as a function of the
average of those other two ($a$). 
This is to connect with the topic of epitaxial
films that are grown on a square
substrate, a common situation for perovskites grown on perovskites or on 
other materials of cubic symmetry---the bulk forms are likely to adapt
more easily to the substrate in the orientation that matches best two of its
lattice parameters to the substrate square parameter.
The points in that figure can be divided in three groups. 
In most cases, $c/a$ is between 0.9 and 1.1, and we find there 
typical phases
of perovskites, such as the $R3c$, $C2/c$, and $Pnma$ ones.
When $c/a$ is larger than 1.1 we obtain the supertetragonal phases
similar to the ground state of BiCoO$_3$ that we have already mentioned.
There is a third set of points, with $c/a$ below 
0.9, corresponding to an orthorhombic $Pmc2_1$ phase first reported
by Yang {\em et al.}\cite{Yang2012PRL} after 
they found computationally as favorable epitaxial phase in the region of large
tensile strains of BiFeO$_3$ and PbTiO$_3$ (they quote a strain of 5\%).

Now we show that this phase is a local minimum {\em of the bulk}
of BiFeO$_3$, and a saddle point in some of the related materials.
This is relevant because it implies that, like for its supertetragonal
phase, BiFeO$_3$ can stand nominally large
epitaxial strains without creating a myriad of defects---the material grows
into a phase whose lattice parameters match very well those of the substrate,
and therefore very small stresses are present in the film.

\begin{figure}
\subfigure[]{
\includegraphics[width=30mm, angle=-90]{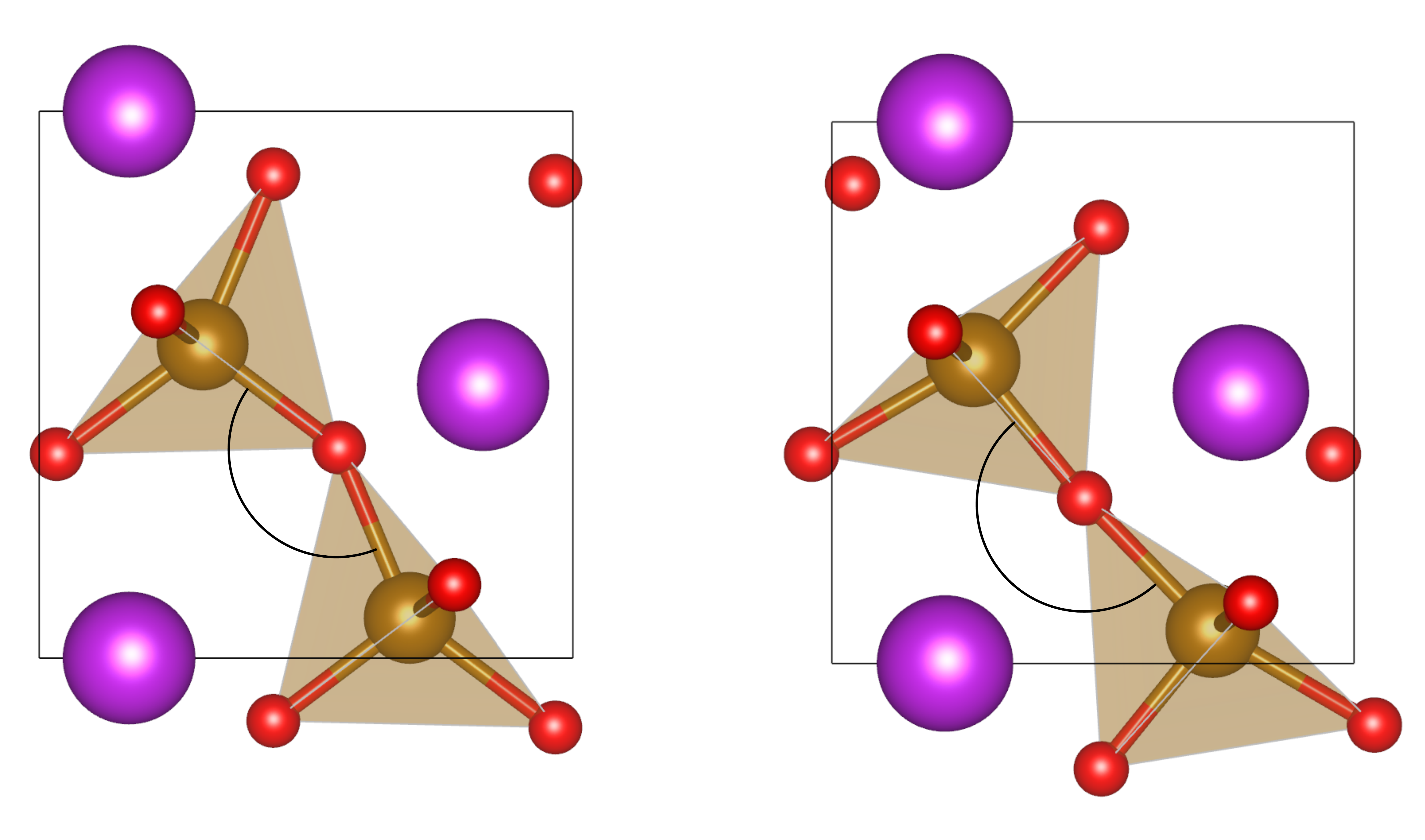}
}
\subfigure[]{
\includegraphics[width=70mm]{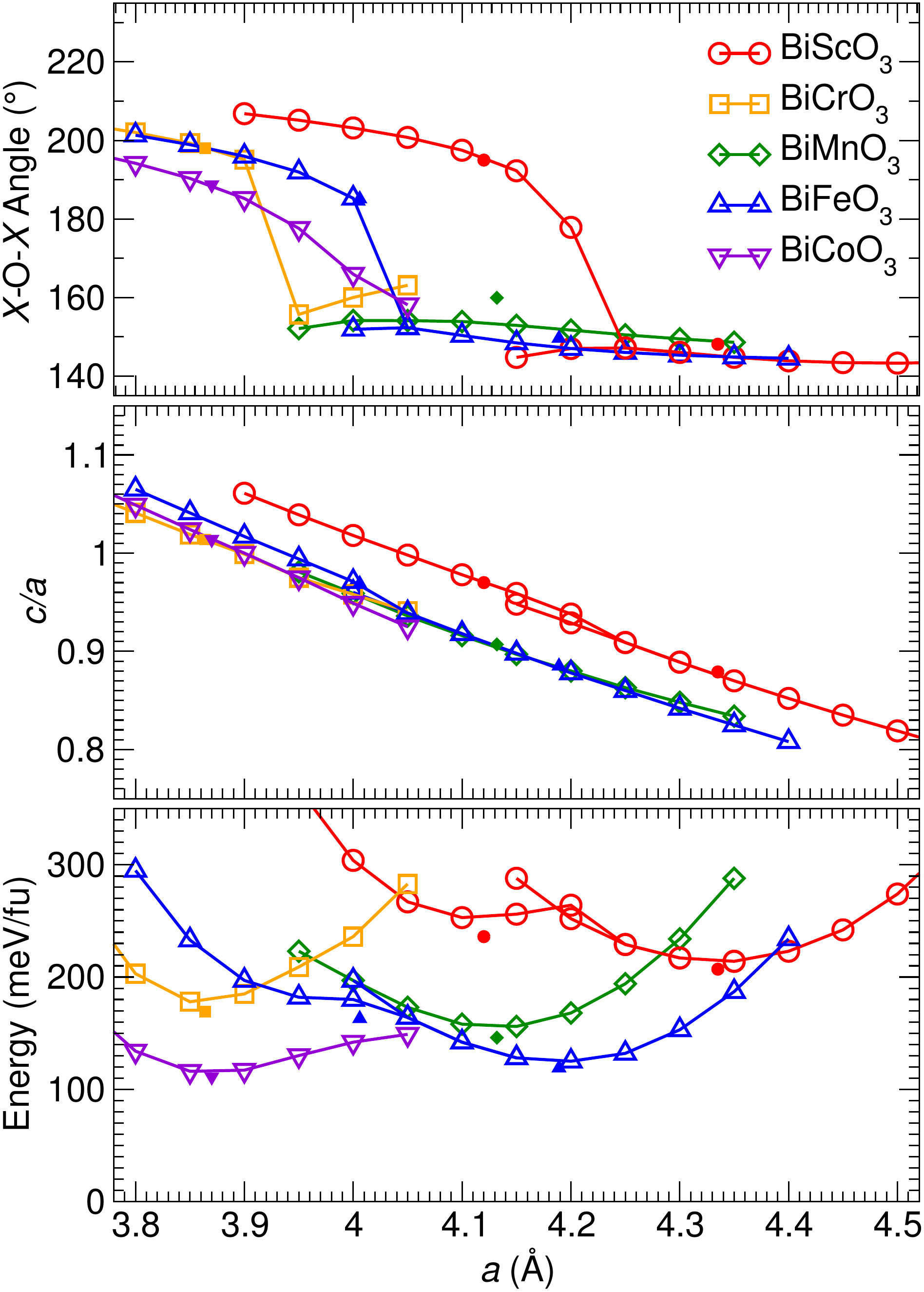}
}
\caption{
(a) Primitive 10-atom unit cell of the inverse supertetragonal phase of
$Pmc2_1$ symmetry with $c/a$ ratio much smaller than 1 (left), and of
the related phase that has $c/a$ around 1 (right); the $X$--O--$X$ angle
that distinguishes these phases is highlighted.
(b) Behavior of this $X$--O--$X$ angle, $c/a$, and energy relative to the 
ground
state for films of those two phases when grown epitaxially on a square
substrate of given in-plane lattice parameter $a$;
the small solid symbols
represent bulk properties of the phases labelled as $Pmc2_1$ in 
Table~\ref{tab_bixo3}.
}
\label{fig_tensile}
\end{figure}

There is an interesting detail in Table~\ref{tab_bixo3} related to this
inverse supertetragonal phase (initially found in our BiFeO$_3$
{\sc uspex} search): 
another phase with the same
set of main distortions and the same $Pmc2_1$ space group exists, but
with a $c/a$ ratio much closer to 1.
This similar phase appeared in the {\sc uspex} search
of BiCoO$_3$ (111 eV/fu above the $P4mm$ ground state).
Apart from $c/a$, the difference between the two phases is that the
$(a^0 a^0 c^+)$ rotation occurs in opposite directions
with respect to the mix of polar and antipolar Bi 
displacements, as shown in Fig.~\ref{fig_tensile} (a).
As seen in the Table, both phases are present in BiFeO$_3$ and BiScO$_3$,
but we have only confirmed one of them for the other three materials.

With the goal of further understanding these phases and of assesing their
suitability as possible ground states in epitaxial films at large
tensile strains, Fig.~\ref{fig_tensile}(b) shows the results of
constraining them to square substrates.
In particular, the bottom graph shows that it costs 
little energy to adapt the bulk phases to the elastic constraint imposed by
the substrate
(the energy difference between the minimum of the open symbols curves and the
small filled symbols is very small). 
It also reveals that these two isosymmetric structures correspond
to two actual different phases, as it can be seen more clearly for the cases
of BiFeO$_3$ and BiScO$_3$, where at intermediate values of $a$ the
two types can be optimized.
The middle graph shows how the $c/a$ ratio is almost linear with $a$, and
almost identical for all cases, but when the larger Sc$^{+3}$ ion is present.
The top graph gives the clearest evidence of the values of $a$ for which
one or the other phase is the favored one.
Finally, the energy curves help to understand why in some cases our bulk
calculations
find only one of the two phases: they are very close in the search space 
of crystal configurations, and when free bulk
optimizations are performed these end up in the lowest special point
available---the inverse supertetragonal one (BiMnO$_3$) or its companion
(BiCoO$_3$ and BiCrO$_3$).
For BiFeO$_3$ and BiScO$_3$ the two phases are further away from each other,
and optimizations can access either of them.

In all, our calculations related to this inverse supertetragonal phase
reinforce the point made by Yang {\em et al.}\cite{Yang2012PRL} that
materials such as BiFeO$_3$ (and perhaps other Bi-based perovskites if they
can be
stabilized) might exist in epitaxial polymorphs with a large in-plane
polarization, and with other interesting properties that arise due to the
special network of square pyramids present.

\subsection{Band Gaps}

In recent times, the search for clean energy has spurred the optimization
of materials for converting solar light into electricity.
Optimal materials have bandgaps that are smaller than those typical
of ferroelectric perovskite oxides, but efforts have been made to reduce
those bandgaps, and in this way couple the ferroelectric functionalities
to those of an energy material.\cite{Kreisel2012NM,Grinberg2013N}
This motivated us to analyze the bandgaps predicted for the phases
identified in our search.

\begin{figure}
\includegraphics[width=60mm]{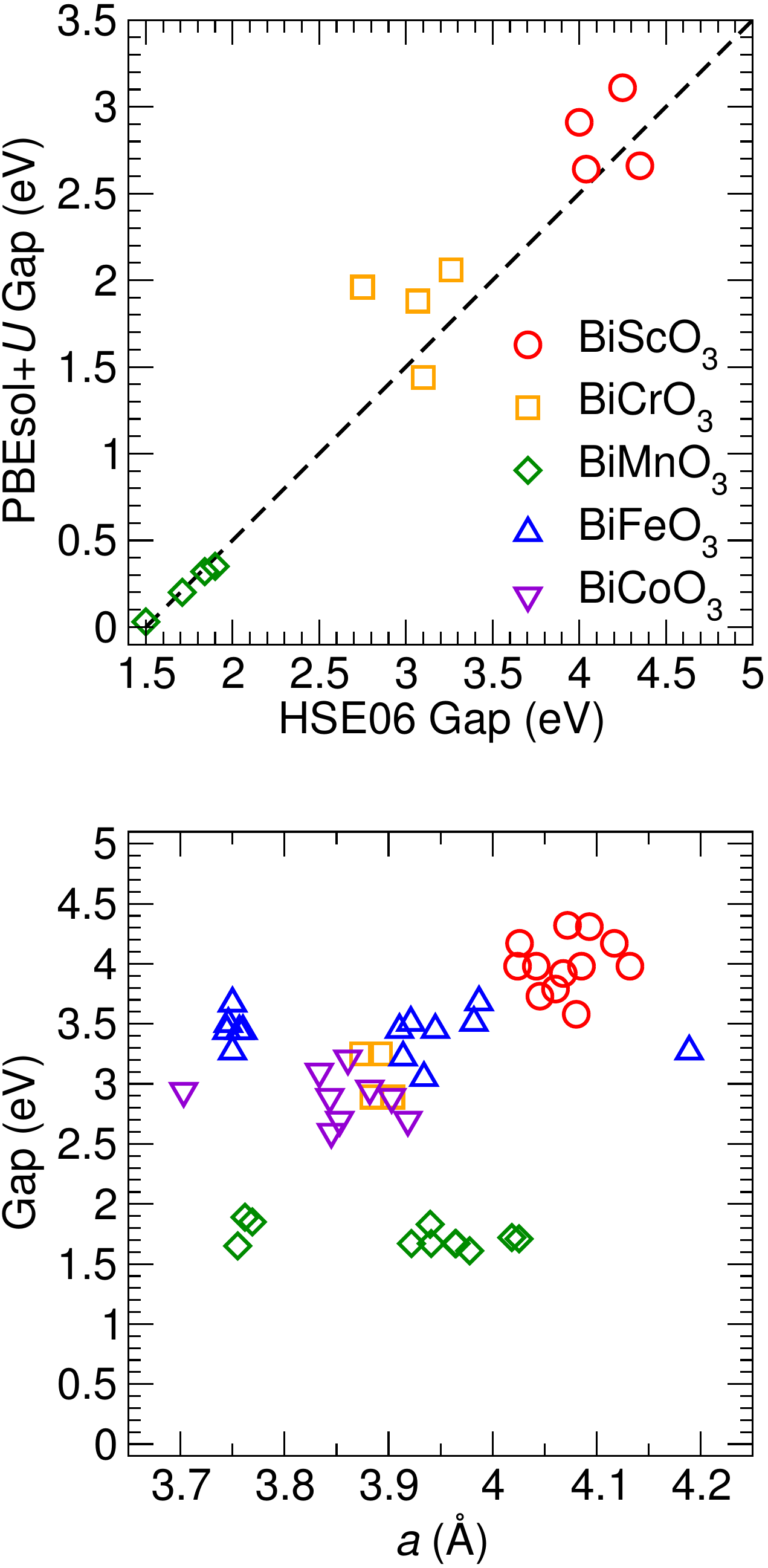}
\caption{
Top panel: comparison between band gaps computed with DFT+$U$ and with HSE06;
the discontinuous line shows that the predictions by the first approach
is systematically lower than the one by the second approach, by a shift
of around 1.5 eV.
Bottom panel: value of DFT+$U$ bandgap for all phases reported in 
Table~\ref{tab_bixo3} that are energy minima
(shifted by 1.5 eV to reflect the most accurate hybrid
calculations)
as a function of the average between the two 
closest lattice parameters of their pseudocubic unit cell.
}
\label{fig_gap}
\end{figure}

Experimentally, the optical bandgap of BiFeO$_3$ has been measured by several
groups, with results ranging from 2 to 3 eV.\cite{Catalan2009AM}
For the other Bi-based materials of this study much less is known about the
value of their bandgap.
McLeod {\em et al.}\cite{McLeod2010PRB} estimated them from 
soft X-ray emission and X-ray absorption spectroscopy measurements, 
and obtained values between 0.9 eV (BiMnO$_3$ and BiFeO$_3$) and 2.6 eV
(BiScO$_3$)---the value for BiFeO$_3$ is significantly lower than what others
have measured.
Computationally, it is well known that standard DFT implementations predict
bandgaps that are systematically smaller than experimentally reported ones,
and that hybrid implementations such as HSE06
produce much better estimations of the gap.\cite{Krakau2006JCP}
Because of this, we show in Fig.~\ref{fig_gap} (top)
a comparison between
our DFT+$U$ methodology and our HSE06 methodology; this shows that the 
former predicts bandgaps that are around 1.5 eV lower than the latter.
With this in mind, Fig.~\ref{fig_gap} (bottom) shows our DFT+$U$ 
computations of the band gap of the phases of Table \ref{tab_bixo3}
(shifted by 1.5 eV, and
done in a reciprocal space grid that is twice more dense in
each direction that the one used to optimize the structures).
From these results we conclude that,
for a given composition, the band gap is not strongly dependent on the
particular polymorph. Further, BiMnO$_3$ clearly appears as
the most promising material 
in this family for photovoltaic applications.

\subsection{Why So Many Polymorphs?}

As mentioned earlier, when similar {\sc upsex} phase
searches are applied to prototype 
ferroelectric perovskites BaTiO$_3$ and PbTiO$_3$ and to Bi-based perovskites
Bi$X$O$_3$, few minima appear in the former cases and many minima appear in the
latter one.
How can we understand the difference between these crystals
based on the characteristics of the $A$ and $B$
cations?

One particularity of the Bi-based perovskites is the lone pair of
$6s$ electrons of Bi$^{3+}$ that overlaps with the O orbitals
to attain a lobe-shape form, as shown by, e.g., electron-localization function
analyses.\cite{Seshadri2001CM,Ravindran2006PRB,Dieguez2011PRB}
The presence of these lobes breaks the spherical symmetry of the Bi$^{3+}$ 
ions, and it causes them to move out of high-symmetry positions---this is
how ferroelectricity originates in BiFeO$_3$ and related materials.
But those lobes can be accommodated in many other ways (paying small energy
prices) that give rise to local minima of the energy illustrated in 
Table~\ref{tab_bifeo3} for BiFeO$_3$.

Ba is an spherical ion, and therefore in BaTiO$_3$ it will not add variety to 
the structures of this perovskite.
However, Pb does have a $6s$ electron pair, so
the case of PbTiO$_3$ must be analyzed further.
There are two main differences between Bi and Pb:
(i) Bi$^{3+}$ is smaller (Shannon radius of 1.17~\AA~at maximum
coordination reported\cite{Shannon1976AC}) than 
Pb$^{2+}$ (Shannon radius of 1.49~\AA);
and (ii) Bi$^{3+}$ is more electronegative than Pb$^{2+}$
(2.02 versus 1.87 in the Pauling scale\cite{Allred1961JINC}).
The $s$/$p$ mixing leading to the active lone pair arises from the 
interaction with the $p$ orbitals of oxygen; since Bi$^{3+}$ is more
electronegative there should be a better match in this respect, and since 
Bi$^{3+}$ is smaller it should adapt better to the electronic requirements
of the metal-oxygen lattice to distort.
In any case, it is an oversimplification to assume that the energy surface of
PbTiO$_3$ has one unique feature (its tetragonal minimum);
for example, it is known that its cubic phase has tilt instabilities,
which only decay into an untilted phase as strain coupling
occurs.\cite{Ghosez1999PRB,Wojdel2013JPCM}

The differences between BaTiO$_3$, PbTiO$_3$, and Bi$X$O$_3$ can be
therefore attributed mainly
to the uniqueness of Bi$^{+3}$'s lone pair, but the properties of the
transition-metal cation
also play a role.
In particular, they must be responsible for
the variations in the structures of the Bi$X$O$_3$ oxides, where 
the formal $d$ electron count is the main difference between transition-metal
atoms that in some cases are very similar in size (the Shannon 
radii\cite{Shannon1976AC} of Sc$^{3+}$, Cr$^{3+}$, Mn$^{3+}$, Fe$^{3+}$,
and Co$^{3+}$ are 0.745~\AA, 0.615~\AA, 0.645~\AA, 0.645~\AA, and 0.61~\AA,
respectively).

For a $d^0$ perovskite (BaTiO$_3$, PbTiO$_3$, BiScO$_3$) 
distortions away for the ideal cubic structure can stabilize the system because
only the metal-oxygen bonding levels are filled, and none of the metal-based
orbitals (which are metal-oxygen antibonding) are occupied.
In BaTiO$_3$ atoms move most favorably towards a triangular face of a O$_6$
octahedron, resulting in its $R3m$ rhombohedral phase.
The presence of a lone pair and a large Pb$^{+2}$ cation changes this balance
and makes a tetragonal phase more stable in PbTiO$_3$.
A smaller ion as Bi$^{+3}$ results, as mentioned above, in more freedom
for the ion to explore its surroundings, and consequently in more phases.

Once the $d$ electron count increases, more possibilities open.
For example, vanadates ($d^1$) have a tendency to exhibit a square pyramidal
coordination,\cite{Zavalij1999AC}
and they do so in a perovskite like PbVO$_3$,\cite{Belik2005CM}
giving rise to a supertetragonal structure.
BiCoO$_3$ assumes a related high-spin d$^6$ configuration (a spherically
symmetric high-spin $d^5$ configuration plus a $d^1$ one) and therefore 
exhibits the same type of $P4mm$ structure in its ground state.
Now of course the other cation is Bi$^{3+}$, and thus many more possibilities
arise.

Systems like BiScO$_3$ ($d^0$), BiFeO$_3$ (high spin $d^5$), and even
BiCrO$_3$ (high spin $d^3$, occupying all $t_{2g}$ levels with one electron) 
contain a transition-metal atom with some sort of electronic closed shell,
so in principle they should be quite similar; indeed
we see some of the structures in Table~\ref{tab_bixo3} being shared for these
materials. 
However, finer effects are also at play, involving O$_6$ rotations
that are mainly combinations of three types: $(a^- a^- a^0)$,
$(a^0 a^0 c^+)$, and $(a^0 a^0 c^-)$.
These add variety to the structures, and result in phases like $C2/c$ being
also quite low in energy.
This is because $C2/c$ combines the very energetically favourable 
$(a^- a^- c^+)$ rotations with parallel and antiparallel Bi displacements to
accommodate the lone pair lobes.
In this respect, it represents a compromise between the paraelectric
$Pnma$ phase and the strongly polar $R3c$
phase.

The Jahn-Teller active ions Mn$^{+3}$ ($d^4$) and Co$^{+3}$ ($d^6$) add
extra possible distortions, reflected for example in the breaking of symmetry
that affects the phase that is the ground state of $R3c$ and energetically
competitive in BiCoO$_3$ and BiMnO$_3$, but with space group $Cc$.


\section{Conclusions}

In this article we report tens of metastable structures of Bi$X$O$_3$ 
($X={\rm Sc}, {\rm Cr}, {\rm Mn}, {\rm Fe}, {\rm Co}$) compounds.
These are minima of the energy within 200 meV/fu or less of the ground 
state, as computed using methods based on DFT (more minima exist at higher
energy differences).
This large degree of polymorphism is related in part to the same mechanism that
is responsible for large values of the polarization in many of them:
the lone electron pair of the Bi$^{3+}$ ion.\cite{Seshadri2001CM,
Ravindran2006PRB,Dieguez2011PRB}
These materials can accommodate those pairs in many combinations of polar
and antipolar displacements that result in many energy minima.
In addition to this, three other factors add variety to these structures:
(1) several available O$_6$ rotation patterns, mainly based on
combinations of $(a^- a^- c^0)$, $(a^0 a^0 c^+)$, and $(a^0 a^0 c^-)$;
(2) the possibility of breaking one of the $X$--O bonds to create stable
supertetragonal phases in BiMnO$_3$, BiFeO$_3$, and BiCoO$_3$, and inverse
supertetragonal phases in BiFeO$_3$; and (3) Jahn-Teller distortions in
BiMnO$_3$ and BiCoO$_3$.

In this way, we see that the variety of lowest-energy states in this family 
is just a reflection of the fine balance between the energies involved in
the cation displacements and the octahedra rotations.
Our calculations show that the reported experimental structures of these
materials actually exist in all of them (with the exception of the variations
of the
supertetragonal structure of BiCoO$_3$, which do not exist in BiScO$_3$
and BiCrO$_3$ as minima).
When the materials are grown, the small energy differences involved
translate into the dramatic structural differences seen.
Epitaxial strain and pressure may be used to shift this balance
and favor different polymorphs (including polar ones), as 
it has been done already in
supertetragonal BiFeO$_3$.\cite{Bea2009PRL,Zeches2009S}

While many calculations based on DFT agree that the ground states of
BiCoO$_3$ and BiFeO$_3$ are the ones found experimentally,\cite{} 
the situation
for BiScO$_3$, BiCrO$_3$, and BiMnO$_3$ is not so simple.
Our calculations using the accurate HSE06 hybrid functional show that the 
experimentally reported $C2/c$ phase is in competition with a similar
phase where the primitive unit cell doubles and an extra $(a^0 a^0 c^+)$ O$_6$
rotation appears.
In the case of BiCrO$_3$, this new phase is slightly lower in energy, and so
are the $R3c$ and $Pnma$ phases.

As part of this research, we have also identified local
minima of the energy in BiMnO$_3$ where charge ordering is present---Mn$^{2+}$
and Mn$^{+4}$ ions alternate in a rocksalt pattern inside the O$_6$ 
octahedra.
These phases are not far in energy from the ground state, and they might
be stabilized by growing BiMnO$_3$ on double perovskites that favor this charge
ordering.

A detailed study of the $Pmc2_1$ inverse supertetragonal phase found here
as a bulk minimum adds to the evidence that BiFeO$_3$ might
grow into this polymorph on square-symmetry substrates with in-plane
lattice parameter around 4.2~\AA;
this was proposed earlier by 
Yang {\em et al.},\cite{Yang2012PRL} 
who identified this polymorph as a favored configuration
in epitaxial films under large tensile strains.
In other materials of the family this is also a special point of the energy
surface, and therefore amenable to stabilization.

Finally, calculations of band gaps for these structures show that BiMnO$_3$
is the most promising of these materials for optical applications, since
it has the lowest band gap of the family---between 1.5~eV and 2.0~eV 
according to our HSE06 hybrid calculations.


\section*{Acknowledgements}
{
O.D.\ acknowledges funding from the Israel Science Foundation through
Grants 1814/14 and 2143/14.
Work in Bellaterra was supported by MINECO (Spain) through Grant
FIS2015-64886-C5-3-P as well as the Severo Ochoa Centers of Excellence Programs
under Grant SEV-2015-0496, and by Generalitat de Catalunya (2017SGR1506).
J.I.\ was supported by the Luxembourg National Research Fund through Grant
No. P12/4853155 COFERMAT.
}


\end{document}